\documentstyle[aps,epsf]{revtex}

\newcommand{\kbt}{k_BT}
\input{epsf}
\input{psfig}
\newcommand{\be}{\begin{equation}}
\newcommand{\ee}{\end{equation}}
\newcommand{\eqn}[1]{Eq.~(\ref{#1})}
\newcommand{\fig}[1]{Fig.~(\ref{#1})}
\begin{document}
\title{ Step-Bunching Transitions on Vicinal Surfaces and Quantum N-mers }
\author{V.~B.~Shenoy$^1$, Shiwei Zhang$^2$ and W.~F.~Saam$^3$\\
{\small $^1$Division of Engineering, Brown University, Providence, RI
02912.}\\
{\small $^2$ Department of Physics and Department of
Applied
Science, College of William and Mary, Williamsburg, VA 23187.}\\
{\small $^3$ Physics Department, The Ohio State University, Columbus,
OH 43210.}
}
\date{\today}
\maketitle
\begin{abstract}
We study vicinal crystal surfaces within the terrace-step-kink model on a discrete
lattice. Including both a short-ranged attractive interaction and a long-ranged
repulsive interaction arising from elastic forces, we discover a series of phases
in which steps coalesce into bunches of $n_b$ steps each. The value of $n_b$ varies
with temperature and the ratio of short to long range interaction strengths. For
bunches with large number of steps, we show that, at $T=0$, our bunch phases
correspond to the well known periodic groove structure first predicted by
Marchenko. An extension to $T>0$ is developed. We propose that the bunch phases
have been observed in very recent experiments on Si surfaces, and further
experiments are suggested. Within the context of a mapping of the model to a system
of bosons on a 1D lattice, the bunch phases appear as quantum n-mers.
\end{abstract}

\section{Introduction}

 The study of the equilibrium properties of a stepped vicinal crystal
surface is important from technological and fundamental perspectives. The
understanding of the surface morphology is key to phenomena like epitaxial growth,
chemical etching and catalysis. In addition, it also provides a
fascinating example of a problem which has a much broader context,
reaching, via a mapping onto a one-dimensional quantum chain, into the realm of 1D
quantum liquids. In this paper we consider the morphology of vicinal surfaces
in the case where step-step interactions are repulsive at large
separations and attractive at short distances. Our work is aimed at
explaining the features of the surface morphology of miscut Si (113) surfaces
observed in two sets of experiments. The first is that  that of the description of
apparent tricritical phenomena observed in a beautiful set of experiments by Song
{\it et al.} \cite{sm1,sm2,sm3} on silicon surfaces oriented between the (113) and
(114) crystalline directions. The second is the very recent observations of
multiple-height steps by Sudoh { \em et al.}\cite{sudoh} on vicinal (113) surfaces
misoriented towards a low symmetry azimuth which is directed $33^0$ away from the
$[\bar{1}10]$ direction towards the $[\bar{3}\bar{3}2]$ direction.    

Recent theoretical work\cite{lassig,bhat} attempted to
explain the experiments of Song {\em et al.} in terms of a continuum model of steps interacting
via
a long-ranged repulsive elastic interaction and a short-ranged
attractive
interaction. This model produces a tricritical point but requires
somewhat
artificial tuning to give the observed coexistence curve exponent
and fails to describe the observed bunching of steps on the vicinal
surfaces coexisting with the (113) facet. Furthermore,
the experimental system seems to be very close to or in the region
where the renormalization group method that is used to study the
model fails. 
Here we explore the
consequences
of retaining the discrete, atomic nature of steps on a crystal surface
within the same model. We discover entirely different physics. The steps
do not phase separate but instead can coalesce into bunches whose size
depends on the relative strengths of the short- and long-range
interactions and temperature. Vicinal surface phases can be characterized by widely
separated n-bunches and transitions occur between phases having
different
values of n. We propose that bunch phases with large n correspond to the
two-phase coexistence region of Song {\it et al} and that the n=2,3,4
bunch
phases produce the multiple-height steps seen by Sudoh {\it et al}.
 For bunches with large number of steps, we show that, at $T=0$,
our bunch phases correspond
to the well known periodic groove structure predicted by Marchenko\cite{mar2}. While Marchenko's calculations are valid only at $T=0$, our theory
predicts the evolution of the periodic groove structure as the
temperature
of the surface is raised. In addition, we show that 
Marchenko's phenomenological
parameters like the edge energy can be determined in terms of the parameters that
characterize the
long and short range step-step interaction strengths.   
  The physics of  bunching transitions that we study here is very
different from the recently studied  step bunching  due
to applied stress, electro-migration and other kinetic effects\cite{dnv}.
In our case, the bunching transitions occur in an equilibrium setting 
due to competing interactions acting on different length scales.
The relevance of
the bunching transitions described here go well beyond
the crystal surface problem. In particular, the model provides
opportunities to study many-body phenomenon arising from coupling of
modes on different length and energy scales. A brief description of
this work has already been reported\cite{vzs}.

The model we use is the terrace step kink model (TSK)\cite{jrs}. This lattice model
can be mapped onto an equivalent quantum mechanical model of interacting spin-less
fermions or hard-core bosons. Details are found in earlier work\cite{jrs,deg,bk}. 
 In the quantum picture it is well-known that the
1-bunches form a Luttinger lattice liquid. Our results generalize this
picture to a Luttinger lattice liquid which can form dimers (the 2-bunches),
trimers (the 3-bunches), and in general n-mers. The transitions between these
phases are quantum many-body phenomena of a type hitherto unexplored. 
In the following section, we introduce the TSK model and develop the
mapping to the hard core boson problem. In Section III, we analyze the model in 
what we call the extreme dilute limit (EDL), where bunches are widely separated.
For $T=0$,
the exact solution clearly reveals the
essence of our
problem, a series of first-order bunching transitions. For bunches
with large numbers of steps, we show the
connection between the bunches phases and the periodic
groove structure predicted by Marchenko\cite{mar2}. In the same EDL,
using exact
diagonalization on small systems to reveal the nature of the bunching
transitions, we
develop the physics for $T>0$. Section IV uses phenomenology coupled
with scaling
arguments to include interactions between bunches in the case that they
are widely
separated and to study the effects of the interactions on transitions
from one
bunch size to another. 
In Section V
we use Quantum Monte Carlo simulations to check the results of exact
diagonalization by
explicitly looking at particle-particle correlations through a series
of bunching
transitions. Section VI applies our results to crystal shapes.
Section VII gives a detailed comparison of our results with
continuum
theories. In Sec.VIII we compare the key
experimental findings with the predictions of 
the theory.  Differences and
similarities
are highlighted, and directions for further development of the theory
are suggested. We conclude by discussing the main results and pointing
directions of future research. 

\section{Hamiltonian of Interacting Steps}

In order to study the thermodynamics of surfaces that are miscut
by a small amount from a reference surface, we will first derive the
surface free energy by treating the miscut angle and temperature
as
thermodynamic variables. The surface free energy will be expressed 
in terms of the ground state energy of a one-dimensional
quantum mechanical model that will be derived using transfer matrix
methods. 
 In the study of interacting steps it is common to use both discrete
and continuum models. In this section we introduce  the
TSK lattice model and the equivalent
quantum mechanical model. We do not go into all the details of the
derivation, but refer the reader to earlier work\cite{jrs}.
We then compare the
kink energy in the lattice model to the step-stiffness in the
continuum model. It turns out that it is important
to understand the limiting process which
brings out the equivalence of the two models.
While it is generally true that when the
limiting process is carried out properly, it is
possible to switch between the models, caution should
be exercised against making erroneous
conclusions about one model based on the results of the
other. The bunching transition is one such example, where
the cut off distance (lattice spacing) in the lattice model
plays a crucial role.  As we discuss in Sec.~VII, there is
nothing corresponding to bunching transitions in the
continuum model; they are purely lattice effects.
On the other hand,  to study the interactions between steps
whose average separation is much larger the lattice spacing,
one can use either the lattice model or the continuum model.
We will now turn our attention to the TSK model and (1) derive the
quantum hamiltonian in terms of hard-core boson
operators, (2) make the connection between the free energy of the
surface to the ground state energy of the quantum mechanical system
and (3) derive the connection between the kink energy in the lattice
model
and step stiffness in the continuum model.

\subsection{Terrace Step Kink Model}

 Let us consider a square lattice with $M$ rows and $L$ columns as shown
in
\fig{steps}, where steps
between terraces are specified by bonds on this lattice.
Let the steps in the TSK model run parallel to the
 $y$-axis, so that the slope of the crystal face is in the
$x$-direction. One can specify a configuration of this system by
specifying the $x$-coordinates of each of the $N$ steps
at each of the $y$-coordinates. The lattice spacing in the
$x$ and $y$ direction is denoted by $a_x$ and $a_y$
respectively. The density of steps is related to the
slope of the interface under consideration through the relation
$s = tan(\theta) = (Na_z/La_x)$ where $a_z$ is the
lattice parameter in the $z$-direction or the step height.
For notational convenience we specify a
configuration  by the state vector
$|x_1,x_2,...,x_N\rangle_k$,
where $x_j$ denotes the position of the $j$th step in the $k$th row.
 The position of the step can only change by $\pm 1$ between the
rows by creating a kink.
 The transfer matrix describes the propagation of steps from row to
row. It can be determined by a kink energy $\Gamma$, an energy
$\eta_0$
for a unit length of the straight step, and an interaction energy
$V_{ij}$
between  parallel straight pieces of steps labeled $i$
and $j$. To render interactions between different rows (steps
interacting
at different values of $y$) tractable, we treat them in an average
way: we compute $V_{ij}$ as the interaction between a unit length
of step at $i$ and an infinite straight step at $j$. We expect this
approximation to preserve the qualitative physics of interest here.
In going from row $k$ to row $k+1$, if the ordinates of
$p$ steps are unchanged while the other $(N-p)$ steps change by
$\pm 1$, then the transfer matrix element has a contribution
$\exp \left(-\beta a_y N \eta_0-\beta \Gamma(N-p)\right)$, 
where $\beta=1/k_BT$.
 The matrix element vanishes
if $x_i=x_j$ for $i\ne j$, i.e. the steps do not cross. The transfer
matrix element also has a multiplicative element contribution from the
step interactions, that can be written as $\exp\left(-\beta a_y\sum_{i
< j}V_{ij}\right)$.
A convenient operator representation is obtained by associating a step
creation operator $a_j^{\dagger}$ for a step at $x_j$ in row $k$:
$a_j^{\dagger}a_j$ denotes a straight step while $a_{j+1}^{\dagger}a_j$
and $a_j^{\dagger}a_{j+1}$ denote a kink to the right and to the left,
respectively. The $a$'s are defined so that $(a^{\dagger}_j)^2 = 0$,
which imposes the non-crossing condition. This gives us a choice of a
fermionic
or hard-core bosonic representation of $a$'s. We adopt the
hard-bosonic
representation, where $a$'s satisfy
 the usual bosonic commutation relations
\begin{equation}
\left[a_i,a_j\right] = \left[a_i^{\dagger},a_j^{\dagger}\right] =
\left[a_i,a_j^{\dagger}\right]=0
\end{equation}
when $i \ne j$ and satisfy  anti-commutation relations on the same site,
{\em
i.e.,}
\begin{equation}
\left[a_i,a_i\right]_{-} = \left[a_i^{\dagger},a_i^{\dagger}\right]_{-}
=
0,
\,\,\,
\left[a_i,a_i^{\dagger}\right]_{-}=1.
\ee
The onsite anti-commutation relations take care of the hardcore
condition
$(a^{\dagger})^2=0$.
 This representation is equivalent to the fermion representation
introduced in Ref.~\cite{jrs}.
 The fermion operators that anti-commute on different sites
are related
to the hard core boson operators through the unitary Jordan-Wigner
transformation\cite{frad}.

Let us  denote the ($N\times N$) transfer matrix by $\hat{{\cal T}}$. For
mathematical
convenience we take the continuum limit in the
$y$-direction (i.e. we take the limit of the lattice
spacing $a_y \rightarrow 0$). In this limit to first order in $a_y$
the transfer matrix can be written as
\be
\hat{{\cal T}} = \hat{1} - \frac{a_y}{k_BT}\hat{H},
\label{coh}
\ee
where $\hat{1}$ is the identity operator and the quantum hamiltonian
is defined as
\be
\hat{H} \equiv \eta(T)\sum_{i}n_i  -t\sum_i\left[a_{i+1}^{\dagger}a_i +
a_{i+1}a_i^{\dagger}-2n_i\right] +
\sum_{(i,j)}V_{ij}n_in_j.
\ee
where $(i,j)$ is used to denote a sum over distinct pairs of bosons,
 $n_i = a_i^{\dagger}a_i$, and the free energy of a straight step 
 $\eta(T) = \eta_0 -2t$, where
 the hopping parameter $t$ is
defined as
\be
t = \lim_{a_y \rightarrow 0}
\frac{\exp\left(-\beta\Gamma\right)}{\beta a_y}.
\label{hop}
\ee
It should be noted that the kink energy $\Gamma$ is a function of
$a_y$ and diverges  in the limit of vanishing $a_y$, but the ratio
given by \eqn{hop} remains finite. Note that since $t$ is a
monotonically
increasing function of $T$ which vanishes as $T\rightarrow 0$, $t$
itself serves
as an effective temperature variable.

 The eigenvalues of the transfer matrix
can be related to the projected surface energy of the miscut surface
through the relation\cite{jrs}
\be
f(s)
=\frac{\gamma(s)}{cos(\theta)}=\gamma_0
-\frac{k_BT}{MLa_xa_y}\log\left(Tr\left[\hat{{\cal T}}^M\right]\right),
\ee
where $\gamma_0$ is the surface tension of the
reference surface and $\gamma(s)$ is the free energy per unit area
of the miscut surface
i.e. the surface tension. In the continuum limit where \eqn{coh}
is satisfied, the projected free energy can be related to the ground
state
energy $E_0\left[N,T\right]$ of $\hat{H}$ through
\be
f(s) = \gamma_0 + \frac{E_0\left[N,T\right]}{La_x}.
\ee
This relation can be used to relate the physical properties of the
surface
like the curvature, stiffness etc. to the ground state energy of the
hard-core boson system. We now turn to a discussion of the form of the
step-step
interactions that will enable us discuss the physics of faceting
transitions.

\subsection{Step-Step Interactions}

In this paper we consider step-step interactions
consisting of a repulsive long range part and an attractive
short range part. We write the inverse square
long range part in the form
\begin{equation}
V^{lr}_{i,j} = \frac{G}{|i-j|^2} \,\,\, ,
\label{long}
\end{equation}
and the short range part  as
\begin{equation}
V^{sr}_{i,j} = -\sum_{k=1}^{n_s}U_k\delta_{|i-j|,k} \,\,\, ,
\end{equation}
where $\delta$ is the usual kronecker delta.   The
quantity $G$ has contributions arising from the
elastic interactions as well as from the electronic charge
rearrangements along the steps \cite{jrs}.
The elastic part depends on the elastic constants of the crystal; for
isotropic crystals it has the form\cite{marc,Noz}
\begin{equation}
G=2\frac{(1-\sigma^2)\vec{f}^2}{\pi Ea_x^2},
\label{gmarpar}
\end{equation}
 where  $E$ and $\sigma$ are 
the Young's modulus and Poisson's ratio respectively and  $\vec{f}$ is the strength of the vector surface dipole force at a
step.
In general for non-metallic crystals the
electronic rearrangements along the steps can be ignored, and
 $G>0$. In the case of metallic crystals,
$G$ can have either sign depending on the details of the electronic
charge rearrangement along the steps.  The short range potential
depends
on the details of the step-step interactions on atomic scales
and has to be worked out from a microscopic calculation. There is ample
experimental evidence
\cite{sm1,sm2,sm3,frohn,heyraud} that
this quantity can be attractive, {\em i.e.} $U_k>0$, on a few lattice spacings
from
the step\cite{red}.  The
purpose of this
paper is to explore the phases of  vicinal surfaces and the associated
crystal
shapes in  crystals with $G>0$ and $U_k > 0$. In the  remainder of the
paper we will consider the case $n_s = 1$ and $U_1 \equiv U$, though
in
Sec.~VII
we comment on the
physics of systems with $n_s > 1$.
For convenience, we re-scale the hopping strength to unity, so that
the interaction parameters get scaled to $G \rightarrow g \equiv G/t$
and $U \rightarrow u \equiv U/t$, while the free energy per step
is rescaled as $\hat{\eta}(T) = \eta(T)/t$. From this point on, we will
focus on the
rescaled Hamiltonian ${\cal H}$, written as
\begin{equation}
{\cal H} = {\hat{\eta}}(T)\sum_in_i -\sum_i\left[a_{i+1}^{\dagger}a_i +
a_{i+1}a_i^{\dagger}-2n_i\right]
+ g\sum_{(i,j)}\left[\frac{1}{|i-j|^2}
-\left(\frac{u}{g}\right)\delta_{|i-j|,1}\right]n_in_j.
\label{rsham}
\end{equation}
Note that $g$ scales inversely with the effective temperature variable
$t$.

\subsection{Step Stiffness and Kink Energy}
In the study of the thermodynamics of steps it is very common to use a
continuum
model, where the Hamiltonian of a single step with coordinate $y(x)$ is
written  as
\begin{equation}
H = \frac{\Sigma}{2}\int {\left(\frac{dy}{dx}\right)^2} dx,
\end{equation}
where $\Sigma$ is the step stiffness.
%
For a single step it is instructive to make  the connection between the
kink energy $\Gamma$ in the lattice model and the stiffness $\Sigma$ 
in the
continuum
model. To this end, in Appendix A we derive, within the lattice
model, the free energy of the step. By comparing
the energies of the continuum model and the lattice model, we see that
the the step stiffness can be written as
\begin{equation}
\Sigma = \lim_{a_y \rightarrow 0}
\frac{k_BT}{2a^2_x}\left(\frac{a_y}{\exp{(-\beta
\Gamma})}\right)=\frac{(k_BT)^2}{2a_x^2t}.
\label{gamma}
\end{equation}
It should be noted that the quantity in the brackets assumes a finite
value
in in the limit of vanishing $a_y$.
The above equation is key in understanding the connection between the
discrete and continuum models. We will return to this equation when we
discuss the stiffness of step bunches.

\section{Phase Diagram in the Extreme Dilute Limit}

The form of the interaction that we have considered for the step
interactions
has competing components
 acting at different length scales. As noted before, we will
focus on the case where there is an attractive potential
on the near neighbor site only i.e $n_s=1$ and $U_1 \equiv U$. 
We now show
that a number of features of this many body system can be
inferred from exact diagonalization of small systems in conjunction with
some simple analytical calculations. In the discussions to follow,
we will present simple physical arguments to show that the for
sufficiently attractive interactions, the steps on the
surface rearrange themselves into bunches at low temperatures.
At $T=0$, where the entropic effects are unimportant, the bunch size
is completely determined by the ratio of the strengths of the
attractive and repulsive interactions i.e. $U/G$.
 With increasing temperature,
the steps start peeling off from the bunches in a series of bunching
transitions.
In the extreme
 dilute limit (the details of which will be clear in the
discussions to follow), we obtain a phase diagram that shows these
bunching transitions as function of temperature.

\subsection{Phase diagram at $T = 0$}

Let us now consider the limit $u \rightarrow \infty$ and $g \rightarrow
\infty$ but $u/g$ finite, so that $t\rightarrow 0$, and the hopping part
of the
Hamiltonian
\eqn{rsham} can be completely
ignored. This corresponds to taking the
zero temperature limit  of the vicinal surface.
 In this limit, the energy of the system is obtained by  minimizing
the total potential energy, and depends on the single parameter $U/G$.
 A little reflection reveals that in the  dilute limit,
 the minimum energy configuration for a given $U/G$ consists
of bosons in bunches of size $n_b$ that are well separated from each
other.
In a bunch, the bosons sit next to one
another. For a bunch of size $n_b$, the energy per boson is
given by
\begin{equation}
\epsilon(n_b)\equiv\frac{E_0[N,T=0,n_b]}{N} = {{\eta}}(0)
-U\left[\frac{(n_b-1)}{n_b}\right] +
G\left[\frac{1}{n_b}\sum_{i=1}^{n_b-1}\frac{(n_b-i)}{i^2}\right] +
O((N/L)^2),
\label{ener1}
\end{equation}
where the term $O((N/L)^2)$ comes from bunch-bunch interactions and
is small in the dilute limit. We call the limit in which the bunch
interactions
are completely ignored, the {\em ``extreme dilute limit''}(EDL).
A straightforward minimization of $\epsilon(n_b)$ in the
EDL reveals the phase diagram shown in \fig{t=0pd}, where the
circles
marked on the $U/G$ axis separate bunches that differ in size by one.
 With increasing
$U/G$, the bunch size keeps increasing; for $U/G <1$ the bunch size is
$1$,
for $1 < U/G < 1.5$ the bunch size is 2, for $1.5 < U/G < 1.83$ the
bunch
size is $3$, for $1.83 < U/G < 2.08$ the bunch size is $4$ and so on.
 At the circles marked in \fig{t=0pd} (i.e. at $U/G = 1,1.5,1.83,2.08,...$)
there
is a coexistence of bunches that differ in size by one. From the graphs
of
the energy per boson vs. $U/G$ shown in \fig{ened} for different $n_b$'s,
we see that the slope at the points where these curves intersect are
not the same. The $T=0$ bunching transitions brought about by
changing $U/G$  can then be considered first order.

With increasing $U/G$, the spacing between the
regions that differ in bunch size by one becomes smaller and smaller.
For large $U/G$, \eqn{ener1} becomes
\begin{equation}
\epsilon(n_b)=\epsilon_\infty+\frac{\epsilon_e}{n_b}-\frac{G}
{n_b}\ln n_b+O(\frac{1}{n_b^2}),
\label{enerlim}
\end{equation}
where $\epsilon_\infty={{\eta}}(0)+G\zeta(2)-U$ is the energy per step
in a large bunch,
$\epsilon_e=U-G-GC$
is the energy associated with the ends of the bunch, and $C=0.577$ is
Euler's constant. Minimizing $\epsilon(n_b)$ in \eqn{enerlim} with respect
to $n_b$ leads to a bunch size given by
\be
n_b \approx
0.561\exp(U/G), \,\,\,\,\,\,\,\,\, U/G \gg 1,
\label{largebunch}
\ee
 which indicates that bunch size diverges rapidly with increasing
$U/G$. 

The above simple calculation has shown that the steps split themselves
into bunches in the EDL at $T=0$. In the next
subsection, we study the unbinding of the steps with increasing
temperature in the EDL. First, however, we connect the result
\eqn{largebunch} for large bunches with earlier work of
Marchenko\cite{mar2}.

For the case at hand, if elastic interactions are ignored, steps have only
the short-ranged attractive interaction $-U$. Here there is a
coexistence (see, {\it e.g.} \cite{jrs}) between the flat, step-free facet
and the facet formed when there are steps at each site. The latter
facet is inclined at an angle $\theta_0 = tan^{-1}(a_z/a_x)$ with respect to the former. All facets with
angles $\theta$ such that $0<\theta<\theta_0$ are unstable and will decompose
into coexisting facets with angles $0$ and $\theta_0$. Marchenko\cite{mar2}
showed that when elastic interactions are taken into account, facets with
$0<\theta<\theta_0$ become stable, being formed from an array of grooves as
shown in Fig.~(\ref{marchenko}). Each groove is composed of a step-free portion
followed by a step bunch, in the terminology of this paper. In fact this
``groove" phase looks just like a bunch phase with $n_b$ large. It should
then be possible to connect Marchenko's results with those of \eqn{enerlim}
in the limit, the EDL, where the bunches are widely separated. This
corresponds to the case of small $\theta$. Marchenko also assumes small
$\theta_0$, as will we in making the connection to his work. In terms of the
geometry shown in Fig.~(\ref{marchenko}), we have 

\begin{equation}
n_b=\frac{L_g\sin\theta}{a\sin\theta_0}\approx\frac{L_g\theta}{a_x\theta_0},
\label{nb}
\end{equation}
where $L_g$ is the length of a Marchenko groove and $a =
\sqrt{a_x^2+a_y^2}$. 
The surface energy per unit
area of the grooved surface is

\begin{eqnarray}
\gamma_g(\theta)&=&\frac{\gamma_0 L_2+\gamma(\theta_0) L_1}{L_g}\nonumber \\
         &\approx&\frac{\gamma_0 (L_2
+L_1\cos\theta_0(1+(\epsilon_\infty/\gamma_0a_x))}{L_g}-\frac{G}{L_g}
\ln(\frac{L_g\theta}{a^*\theta_0})\nonumber \\
&\equiv&\gamma_0(\theta)-\frac{G}{L_g}
\ln(\frac{L_g\theta}{a^*\theta_0}),
\label{gammamar}
\end{eqnarray}
where
\begin{equation}
\gamma(\theta_0)=\gamma_0 \cos\theta_0+\frac{\epsilon(n_b)n_b}{a_x},
\end{equation}
and
\begin{equation}
a^*=a_xe^{\epsilon_e/G}.
\label{astar}
\end{equation}
\eqn{gammamar} is precisely Marchenko's\cite{mar2} result in the EDL if $G$
can be identified as
\begin{equation}
G=2\frac{(1-\sigma^2)\vec{F}^2}{\pi E},
\label{gmar}
\end{equation}
in which  $F$ is the surface force acting at the ends of the
bunch. Now, an expression for $G$ was given in \eqn{gmarpar}
in terms of $\vec{f}$, the strength of the vector surface dipole force at a
step. To connect $\vec{F}$ with $\vec{f}$, note that the surface force
density due to an array of steps with density $n(x)$ is given by
\begin{equation}
\vec{\cal F}(x)= \int\vec{f}\delta^{'}(x-x_0)n(x_0)dx_0=
\vec{f}\frac{dn(x)}{dx}. 
\end{equation}
$\vec{\cal F}(x)$ can be integrated to obtain the force $\vec{F}$ on the surface at the edge of a
bunch. Noting that  the step density has a discontinuity of size $n_0$, where
$n_0=(1/a_x)$ is the uniform step density in a bunch, we obtain
\begin{equation}
\vec{F}=\vec{f}\int\frac{dn(x)}{dx}dx=\frac{\vec{f}}{a_x}.
\end{equation}
This result proves the equivalence of \eqn{gmar} to \eqn{gmarpar}, and demonstrates that our theory is in fact a
microscopic version of Marchenko's. In addition to connecting the
surface
force to the strength of the force dipole at the step, \eqn{astar}
gives an microscopic interpretation to the phenomenological
edge energy, $\epsilon_e$, in the Marchenko theory. Explicitly, in terms of the microscopic step interaction strengths,
$\epsilon_e = U - 0.423G$. 

Minimizing $\gamma_g$ with respect to $L_g$ at fixed $\theta$ and
$\theta_0$ gives
\begin{equation}
L_g=\frac{ea^*\theta_0}{\theta},
\end{equation}
which is Marchenko's result in the EDL and is equivalent to
\eqn{largebunch}. At this minimum, we have
\begin{equation}
\gamma_g(\theta)=\gamma_0(\theta)-\frac{G\theta}{ea^*\theta_0}.
\label{gamg1}
\end{equation}

To further understand the grooved configuration, 
we make use of the relations $L_g\sin\theta=L_1\sin\theta_0$ and
$L_g\sin(\theta_0-\theta)=L_2\sin\theta_0$ to show that
\begin{equation}
\gamma_0(\theta)=\cos(\theta)[\gamma_0+\frac{\epsilon_\infty}{a_x}
\frac{\tan(\theta)}{\tan(\theta_0)}].
\end{equation}
The full projected free energy is, for small $\theta$
and $\theta_0$, given by
\begin{equation}
f_g(\theta)=\frac{\gamma_g(\theta)}{\cos\theta}=\gamma_0+
(\frac{\epsilon_\infty}{a_x}-\frac{G}{ea^*})
\frac{\theta}{\theta_0}.
\end{equation}
The contribution to the  projected free energy at $O(\theta^3)$ comes
from bunch-bunch interactions that are ignored in the EDL.
 In the limit of vanishing $\theta$ that
we are considering, the bunches interact via an inverse square
interaction
of strength $Gn_b^2$. A straightforward computation of the bunch
interaction contribution yields
\begin{equation}
f_g(\theta)=\frac{\gamma_g(\theta)}{\cos\theta}=\gamma_0+
(\frac{\epsilon_\infty}{a_x}-\frac{G}{ea^*})
\frac{\theta}{\theta_0}+\frac{G\pi^2}{6ea^*}\left(\frac{\theta^3}{\theta_0^3}\right).
\label{fg10}
\end{equation}
Marchenko's full result replaces \eqn{gamg1} by
\begin{equation}
\gamma_g(\theta)=\gamma_0(\theta)-
\frac{G}{\pi a^{*}}\sin(\pi\frac{\theta}{\theta_0}).
\label{gamg2}
\end{equation}
When expressed as an expansion in $\theta$, up to $O(\theta^3)$, for small $\theta_0$,
\eqn{gamg2} becomes identical to \eqn{fg10},
 once again
confirming that our model is identical to Marchenko's model at
$T=0$. In the following section we derive the finite temperature
extension of \eqn{fg10}.

\subsection{Phase Diagram for $T > 0$}

 At finite temperatures, it becomes favorable to form kinks in the steps
for entropic reasons. This causes the steps within a bunch to
 fluctuate from their mean positions, leading to a decrease in the free
energy. On general physical grounds then, we expect the free energy per
step
in larger bunch sizes to
 decrease less rapidly compared to smaller bunches as steps within
larger
bunches will have their fluctuations more constrained. If this happens,
the
free energies of smaller bunches will fall below those of larger
bunches. We
then expect
larger bunches to rearrange themselves into smaller bunches,
with increasing temperature, until eventually at very
large temperatures only one step is left in the bunch. In this section,
we
show that this indeed happens.

In order to compute the bunch size at a given temperature in the EDL,
it is not necessary to solve the step problem in the thermodynamic
limit, or equivalently the hard core boson for a large number of
bosons confined to very large number of sites.
Instead, computing the energy per boson in a bunch by solving the
bunch problem by exact diagonalization is sufficient. This
simplification can be
understood by noting that bunch-bunch interactions give rise
to  corrections to the ground state energy that are smaller
than the leading bunch energy by a
factor $s^2$, where $s$ is the boson density introduced
earlier. As a result, in the limit that the interaction between the
bunches are ignored, comparing the energy per boson in
bunches of different sizes will determine the stable bunch size.
Even computing the ground state energy of a bunch
confined on a large  number of lattice sites could be difficult
due to demands on computational resources. However,
for studying the transitions from bunch of size $n_b$ to $n_b -1$
$(n_b > 2)$, the bunches are sufficiently bound that a small
number of lattice sites (typically twice to thrice the
bunch size)  is sufficient to get accurate answers. We will
present more details on this in the discussions to follow.

 Following the notation introduced for $T=0$, we write
the energy per boson as
\be
\frac{E_0[N,T,n_b]}{N} =  {\hat{\eta}}(T) + f(n_b,T)  +
O((N/L)^2),
\ee
where in the EDL, the bunch size $n_b$ is obtained by minimizing
 $f(n_b,T)$, which is the contribution to the
energy arising from interactions within a bunch. It is
obtained by solving for the ground state energy of
Hamiltonian given in \eqn{rsham} for $n_b$ bosons.  We illustrate the
results of this
 procedure
in \fig{eneps} for $u/g = 1.65$. Here the energy per boson
is plotted as a function of inverse temperature $g=G/t$ for
$n_b = 2,3$ and $4$. At low temperatures ($g > 6.8$), the energy
per boson in the
3-bunch the lowest value, while at moderate temperatures ($6.8 > g >
4.54$) the 2-bunch gives lowest energy per boson, while at high
temperatures ($g <
4.54$) the 1-bunch has the lowest energy (the 1-bunch becomes stable
when
$f(2,T) = 0$). From our results we have computed phase
boundaries, shown as the solid curves in \fig{phasd}, for transitions from
1-bunch to
2-bunch, from 2-bunch to 3-bunch, and from 3-bunch to 4-bunch. In
order
to obtain the phase boundaries between the $n_b$-bunch phase and
 the $n_b-1$ phase for a fixed $u/g$, one has to find the value of $g$
that satisfies
the equation $f(n_b,T)-f(n_b-1,T)=0$. We do this numerically
by using a standard root finding algorithm which approaches
the root iteratively\cite{numr}. The root finding algorithm
 requires repeated evaluation of $f(n_b,T)-f(n_b-1,T)$ using exact
diagonalization of the $n_b$ and $n_b-1$ bunches for different
values of $g$.

A key feature of plots shown in \fig{eneps} is that
the energy curves for bunches that differ in size by one
intersect
with a finite slope. This means that the derivative of
the free energy of the vicinal surface  with respect to temperature, or
the
entropy, is discontinuous at the transition, implying that
the the transitions are first order in the EDL. In the
next section we will argue that, when bunch-bunch interactions are
taken
into account, the discontinuity in slope can vanish and the transition
becomes continuous.

In calculating the energy of the 3-bunch and 4-bunch, we confined
the bosons to $12$ sites on a ring, while for the 2-bunch, 200 sites.The 
latter is equivalent to a 1-body problem describing the 
relative motion. These ring sizes are sufficient to get accurate
answers because the bunches are tightly bound with a mean particle
spacing between 1 and 1.5, so that as long as
the ring is bigger than about twice the bunch size,
the energies do not change much. We have also computed the
ground state energies using the GFMC method for bunches confined
on lattices with up to 100 sites and verified that the energies plotted in
\fig{eneps} are accurate. In \fig{exadc}  we show the energies
computed
for a 3-bunch on 12 sites using exact diagonalization
and on 30 and 100 sites using GFMC( see Sec.~VI for details on
GFMC simulations). 
For sake of completeness, we have computed
the mean particle spacing for the 3-bunch along the phase
boundary separating the 2-bunch and 3-bunch regions and for the 4-bunch
along
the phase boundary marked between the 3-bunch and 4-bunch regions (see
\fig{phasd}). These
spacings vary from about 1.3 to 1.6 lattice spacings as $u/g$ is varied
from its
values at the transition onsets up to $u/g=3$. It is clear that the
bunches  are very
tightly bound in all the cases.

 It is of considerable interest to understand how the limit
$g\rightarrow 0$ is
approached for our model. This limit, where only a near-neighbor
interaction
remains, has an exact solution (see, {\it e.g.} Refs.~\cite{jrs,burk})
involving a first-order
phase transition from a gas (infinitely dilute) of 1-bunches to a
condensed phase (an
$\infty$-bunch). 
Note from \fig{phasd} that
the 2-3 boundary
crosses the 1-2 boundary at $u/g=2.4$ so that a 2-bunch is no longer
stable for
$u/g>2.4$. We also show the 1-3 transition line extending from $u/g=2$
upwards. This
line cannot be distinguished from the 2-3 line for $u/g>2.4$, but in
fact lies slightly
below it and above the 1-2 line. Thus, for $u/g>2.4$ in \fig{phasd}
 the 1-bunch goes directly to
the 3-bunch and
then subsequently to the 4-bunch as the temperature 
or $1/g$ is lowered at fixed $u/g$.
Further
computation for values of $u/g$ up to 8, depicted in \fig{phasd2}, reveals
that the 3-4
line comes very close to the 1-3 line while remaining below it. We then
conjecture that
as
$g\rightarrow 0$ the n-1 transition lines become asymptotically equal
for arbitrarily
large n. This is clearly an approach to the 1-$\infty$ transition, which
is the
first-order transition in the pure short-ranged potential case. As the
transition in
the short-ranged case is at $u=2$\cite{jrs,burk}, the asymptotic slope of the
transition lines should
be at $(1/g)/(u/g)=1/2$. This is indeed the case as seen in \fig{phasd2}.

\section{Phase Diagram in the dilute limit}

 We will now consider the effect of the bunch-bunch interactions that
were ignored in the previous section. We adopt the following steps
in discussing this effect: (1) Using the general principles of
quantum mechanics, we argue that when the the inter-bunch interactions
are
turned on, the first order transition should  become continuous.  (2) We
then estimate
the regions around the first order transition lines (obtained
in the previous section) where the inter-bunch interactions become
important.
(3) Finally, we develop a perturbative approach to compute the free
energy to
$O(N/L)^3$ in the region where the inter-bunch interactions are small.
In the next section, we will then present the results of GFMC
simulations on
larger systems and show them to be consistent with our simple theory.
In particular we compute the 2-point correlation function, which
clearly shows the bunching transitions in large systems.

The effects of interactions on the bunches can be understood
qualitatively
using arguments based on general principles of quantum
mechanics. We confine our discussion to those regions in the phase
diagram
where the $n_b$-$n_b-1$ phase boundary is well separated from all 
the other phase boundaries.
Consider \fig{esplit}, where we schematically plot the energy per step in
bunches of size $n_b$ and $n_b-1$ as a function of the inverse effective
temperature $g$ for some fixed value of $u/g$. This is similar to
\fig{eneps}, which was
used to study phase transitions in the extreme dilute limit. In this
figure,
we identify three regimes, namely, (1) regions where the energy per
boson
in the
$n_b$-bunch is much lower than the $n_b-1$-bunch, (2) regions where
the energy per boson in both the $n_b$-bunch and the $n_b-1$-bunch
have comparable values and (3) the regions where the energy of the
$n_b-1$-bunch is much lower than the $n_b$ bunch. In
regions (1) and (3) the ground state energy of a system
with large number of bosons can be estimated using perturbative
 methods. Here to leading order the energy of the system
is the energy of a bunch times the total number of bunches.
The next to leading order comes from the bunch-bunch interactions
via an inverse square interaction. In region (2), called the ``critical
region'', the physics is
more complicated, since the system has the nearly same energy in both
the $n_b$ and $n_b-1$ bunch phases. In this case we expect
a splitting of the energy levels as sketched in \fig{esplit}, where the
degeneracy of the energy levels is shown to be lifted
due to bunch-bunch interactions. The extent of the region with 
significant splitting is denoted by $\Delta
 g(n_b\rightarrow n_b-1)$.
This width, which
vanishes in the limit of vanishing densities, can be written as
\be
\Delta
 g(n_b\rightarrow n_b-1) \sim \Delta T_c(n_b\rightarrow n_b-1) \sim
s^{p_{n_b}},
\label{cexp}
\ee
where $\Delta T_c(n_b \rightarrow n_b-1)$ is the width, in temperature, of
the critical region
 and the exponent $p_{n_b} > 0$ will be computed later in this section
by comparing the free energies of the
bunches of size $n_b$ and $n_b-1$, derived perturbatively.
 The physics in the critical region is
of general interest, and its relevance goes well beyond the problem
we are studying. It involves coupling of modes with very different
 length scales. A tightly bound $n_b$-bunch in region (3) can be
thought of as having phonon excitations involving all the bunches as
well as excitations
within the bunch. While the wavelength of the former type
of excitations is large (of the order of bunch spacing $s^{-1}$), the
latter modes are localized in regions of the order of few lattice
 spacings.
A bunch of size $n_b-1$ in region (1) will also possess similar
 excitations on
both large and small scales. While the couplings between the
excitations on different scales are small in regions
(1) and (3), we expect strong coupling between the short and long
wavelength
modes
in the critical region.

\subsection{Perturbative Treatment for Stable Bunches}

In order to compute the bunch-bunch interaction energy in
regions with stable bunch sizes (like regions (1) and (3) in
\fig{esplit}),
 we make the
following assumptions: (1) If a bunch has $n_b$ steps, then
the stiffness of a bunch is $\Sigma_b = n_b \Sigma$, where $\Sigma$ is
the
single bunch stiffness in the continuum limit. (In the discrete
model, the kink energy required to make a kink in the
bunch becomes $n_b \Gamma$. However, taking the appropriate
limit of vanishing $a_y$ would guarantee that the combination
$\exp(-\beta n_b \Gamma)/a_y$ would scale like $n_b$). In making this
approximation, we have ignored the contribution to the
bunch stiffness that arises from interactions of the steps
in the bunch. This is justified by the fact that average spacing in a
bunch, for $u/g<3$, remains well under two lattice spacings,so that the
steps in
a bunch remain tightly bound as a single
 entity.
The approximation will be less accurate if the step spacing in the
bunch is large, in which case bending of steps would also
alter the interaction energy within the bunch.
(2) The bunches interact with each other with a
 renormalized
inverse square potential of strength $g_b = n^3_bg$.  This
 approximation
is justified by noting that, in the dilute limit, the spacing
between the bunches is very large, so that one bunch
sees other bunches as if they were single lines without
any internal structure.
We note that the contributions to the energy of the system arising from
the short range bunch-bunch interactions are smaller
than the inverse square contribution by a factor proportional
to the density of steps $s \ll 1$. This can be shown
very easily using a Hartree-Fock estimation of the short-ranged
 contribution\cite{hartree}.
The energy of the system can then be computed using the
Calagero-Sutherland model
of hard-core bosons interacting via an inverse square law,
for which an exact solution was provided by Sutherland\cite{Suth}.
Using his
solution, we can write the energy per site in a system with
bunches of size $n_b$ as
\be
\frac{E_0[s,T,n_b]}{L} = {\left[\eta(T) + f(n_b,T)\right]a_x\over a_z}s
+ \left(\frac{{\pi}^2(\kbt)^2
\lambda_b^2 a_x}{6{n_b}^4a_z^3\Sigma }\right)s^3 + O(s^4),
\label{perte}
\ee
where
\be
 \lambda_b =
\frac{1}{2}\left[1+\sqrt{1+2n_b^3g}\right].
\label{lamb}
\ee
Using the above form of energy, one can compute such quantities as
surface stiffness, crystal shapes etc, which we  will turn to
following a discussion of the physics in the critical region.
An important point to make at this juncture is that in the 
limit of zero temperature, for
large $n_b$, \eqn{perte} is identical to  Marchenko's result given in
\eqn{fg10}.  It is then clear that \eqn{perte} is the finite
temperature
extension of Marchenko's result.

\subsection{Analysis of the Critical Region}

In the critical region, since the energies of both the $n_b$ and
$n_b-1$ bunches are nearly equal, we retain the terms of order
$s^3$ arising from bunch-bunch interactions. In order
to estimate the width of the critical region,
we use the criterion  $E(s,T,n_b-1) \approx E(s,T,n_b)$, to obtain
\be
s^2 \sim |f(n_b,T)-f(n_b-1,T)|.
\ee
If
\be
|f(n_b,T)-f(n_b-1,T)| \sim {|\Delta
 g(n_b\rightarrow n_b-1)|}^{\alpha_{n_b}},
\ee
we obtain the result $p_{n_b} = 2/\alpha_{n_b}$, for $p_{n_b}$
introduced
in \eqn{cexp}.
For all the transitions from the 4 to 3 bunches as well as
for the transitions from the 3 to 2 bunches we see that the
curves for the energy per step intersect with a finite slope
for the parameter range that we have studied. This means that
for these first order transitions $\alpha_{4} = \alpha_{3} = 1$, the
width of the critical region scales like
\be
\Delta T_c \propto s^2.
\ee
For the unbinding of the 2-bunch to the 1-bunch, we find,
for $g < 3/2$, $\alpha_2 = 2/\sqrt{1+2g}$, while for $g>3/2$,
$\alpha_2 = 1$, from which one can infer that
\begin{eqnarray}
\Delta T_c(2 \rightarrow 1) &\propto& s^{\sqrt{1+2g}}, \,\,\,\, g < 3/2 \nonumber \\
 &\propto& \,\,\,\,s^2,    \,\,\,\,\,\,\,\,\,\,\,\,\, g \geq 3/2.
\label{cr2}
\end{eqnarray}
As noted earlier the 3 bunch to 1 bunch  transition preempts the
2-bunch to 1-bunch transition for $g < 1.86$. The 3 bunch to 1 bunch
transition was seen to be first order in the entire range
investigated.

Note next that the simplest conjecture for the shift of the
transition temperature due to interactions is that it is of the same
order of magnitude as the level shift given in Eq.~(\ref{cexp}).
Explicitly, for the transitions that are first order in the EDL,
\begin{equation}
|T_c(n_b \rightarrow n_b^{\prime};s) - T_c(n_b \rightarrow
n_b^{\prime};0)| \propto
s^2.
\label{dg}
\end{equation}
While we find that all the transition in the model are first order in
the EDL, it is interesting to compute the shift in transition
temperature for the 2 bunch to 1-bunch transition temperature
when $g < 3/2$. With our conjecture, and using \eqn{cr2} for $g <
3/2$, one obtains
\begin{equation}
 |T_c(2 \rightarrow 1;s) - T_c(2 \rightarrow
1;0)| \propto
s^{\sqrt{1+2g}}.
\label{dg2}
\end{equation}
Note that for this case our conjecture indeed yields the result derived within
 continuum theory\cite{lassig,bhat}  using renormalization group
 methods. 

While we have used simple arguments to predict the width of the
critical region and the shift of the transition temperature, a detailed
analysis of the
nature  of unbinding transitions and the form of the free energy
require further study of the many body problem
involving the thermodynamic limit. One natural way to
study this is by the means of computational methods like
GFMC or exact diagonalization methods for large systems.
In the following section, we discuss the results obtained using
GFMC simulations.

\section{ Quantum Monte Carlo Simulations}
\label{QMCresults}

In the preceding section we argued, on the basis of results for the
{\em extreme\/} dilute limit from exact diagonalization of small
systems, that the bosons undergo bunching transitions as strengths of
the interaction parameters are varied. In this section, we will
perform calculations on larger systems to directly study the dilute
limit and attempt to understand the physics in the ``thermodynamic
limit''.  In addition to exact ground state energy, we also study pair
correlation functions in order to probe the structure of the many-body
state, in particular the occurrence of bunching and transitions
between different bunch sizes.

The primary method we use is Green's function Monte Carlo
(GFMC)\cite{GFMC}, a Monte Carlo method to study ground state
properties of quantum many-body systems which is 
exact for bosons.  Variational Monte
Carlo (VMC) calculations are performed prior to GFMC, in order to
optimize our variational wave function, which is used to generate the
initial state and also as the importance function in GFMC.  In
Appendix II, we present a brief self contained treatment of these
methods.  The algorithm we use follows closely that of Refs
\cite{Triv,Shi}, which we refer the reader to for more details.

The variational many-body wave function used is of 
 the Jastrow form, written as a
product of two body correlations, {\em i.e.,}
\begin{equation}
\psi_T\left(x_1,x_2,.....,x_N\right) = \prod_{(i,j)} f(|x_i-x_j|),
\label{twf}
\end{equation}
where  $|x_i-x_j|$ is the mean  distance between
particle
$i$ and particle $j$. The function $f(r)$, where $r$ is a positive
integer,
is chosen to be of the form
\begin{eqnarray}
f(r) &=& a(r -r_0)^2 + d, \,\,\,\,\,\, r \le r_0 \nonumber \\
&=& 1 - \frac{f_2}{r^{\alpha}} + be^{-b r}, \ \ r > r_0,
\label{varwf}
\end{eqnarray}
Defining $f(1) \equiv f_1$, we choose $ f_2 = (\alpha+r_0)r_0^{\alpha
-1}(1-d)$,
$b = 1 + \alpha/r_0$, $b = \alpha f_2e^{b r_0}/(b
r_0^{\alpha+1})$
and $a = (-d+f_1)/(1-r_0)^2$, thus reducing our variational parameters
to
four, namely, $f_1$, $r_0$, $d$ and $\alpha$. We note that
the parameters are chosen such that $f(r)$ and $df/dr$ at $r=r_0$
are continuous. These parameters have direct
physical implication as shown in \fig{2bcor}.

In both the VMC and GFMC, we
used 500 walkers (on the average for GFMC) for computing the averages.
In the variational calculation, measurements were made
after
10,000 metropolis steps. In the GFMC calculations measurements are made
after 10,0000 steps during which system relaxes from the initial
variational distribution. Measurements were made for
a total of 20,0000 steps, and statistical errors were obtained
by dividing these into 200 blocks of 1000 steps each and estimating
the variance of the block measurements.

 We present results for three separate
cases, namely, $u/g =1.25, 1.65, 2.0$. These cases are chosen so that 
in the limit of $g \rightarrow
\infty$ the ground states have bunches of size $2,3$ and $4$ bosons,
respectively. 

First we show in Table I a comparison of GFMC and exact
diagonalization results for the energy of a small system of 3
bosons. In the exact diagonalization the bosons are confined on 12
sites in a ring, i.e., with periodic boundary condition. The results
are shown together with those from GFMC for the same system. The
agreement is exact. This is to be expected, since GFMC yields exact
ground state energies. Also shown are the energies computed by GFMC
for a ring of size 30 and 100 respectively. We see that they are indistinguishable from the
corresponding results for 12 sites, indicating that at 12 sites the
finite-size effect is already negligible. This confirms our earlier
conclusion that the 3 bosons are tightly bound. In \fig{exadc}, we again
compare GFMC results for 2- and 3-boson systems on 30 sites with our
exact diagonalization results, over a wider range of parameter
space. The excellent agreement reassures that our GFMC code 
is well-behaved, and that the small system sizes used in exact
diagonalization calculations of the EDL are justified.

In table II, III, and IV we present
 the variational wave-function and the energies $E_{VMC}$ and $E_{GFMC}$
and compare them with the approximate energy $E_{approx}$ of the rescaled
Hamiltonian given by Eq.(\ref{rsham}). (For sake of clarity we do not
include the first term which is proportional to the number of bosons,
since it does not directly affect the physics of bunching
transitions.)  For $N$ bosons on $L$ sites, 
$E_{approx}$ is computed from the energy
per boson $f(n_b,T)$ computed using exact diagonalization and 
the approximate interaction energy using 
\begin{equation}
E_{approx} = \frac{Nf(n_b,T)}{n_b} +
\frac{\pi^2\lambda_b^2 N^2}
{3n_b^4L^3},
\label{app} 
\end{equation}
where $\lambda_b$ is given by Eq.(\ref{lamb}). In computing
$E_{approx}$,
the value of $n_b$ is obtained from the
the phase diagram given in \fig{phasd}. For all the values
of $u/g$, for large $g$, we see that there is good agreement
between the $GFMC$ and approximate energies and fair agreement for
small values of $g$. These results confirm the approximations made
in the analysis of bunching and bunch-bunch interactions. 
As $g$ is decreased, the system is 
more quantum mechanical and each bunch is less tightly bound, 
hence $E_{approx}$ becomes worse.
Note 
that the in all cases $E_{GFMC} < E_{approx}$, which is not surprising
 since
we did
not consider the role of the short-range attraction between the
bunches
in evaluating $E_{approx}$.

We also calculate the pair correlation
function
which for $L$ even is defined as
\begin{equation}
g(x) = \frac{L}{2N^2}<\sum_{i \ne i^{\prime}} \delta\left(x-x_i +
x_{i^{\prime}}\right)>,
\label{corf}
\end{equation}
for $1 < x < L/2-1$. The distance is halved due to periodic boundary 
condition.
In the expression
for $g(x)$,  $x_i$ and $x_{i^{\prime}}$ are the coordinates
of the bosons $i$ and ${i^{\prime}}$.  

In GFMC the mixed estimate is used to compute the average indicated in
Eq.~(\ref{corf}).  
Recall that 
the mixed estimate\cite{GFMC} 
for the expectation value of an operator ${\cal O}$ is   
\begin{equation}
\langle {\cal O}\rangle_{mixed} = 
\frac{\langle\psi_T|{\cal O}|\psi_0>}{<\psi_T|\psi_0>},
\end{equation}
where $|\psi_0\rangle$ is the ground state wave function while 
$|\psi_T\rangle$ is the trial wave function.
While exact for the energy, this estimator yields
an approximate ground state expectation value for other
operators. 
If $|\psi_T\rangle$ is good, a simple extrapolation\cite{GFMC} of the
mixed estimate with the variational 
expectation can lead to a
further improved result. 
Our $|\psi_T\rangle$ is relatively poor, giving variational 
estimates of $g(x)$
which are rather structure-less in long range, so we did not 
attempt this extrapolation.
However, as we see below, even
with this $|\psi_T\rangle$, $g(x)$'s computed from 
the mixed estimate clearly shows the bunching
structure. 

For $u/g= 1.25$, \fig{pairc1} shows the pair correlation
function for 10 bosons on 100 sites
respectively. We see that for large $g$ ($g=20$ and
$g=15$), the pair correlation function shows two equal peaks near $x=20$ and
$x=40$. Taking into account periodic boundary condition, we see
that this indicates
the presence of 5 $2$-bunches in the system.
Note that the $2$-bunch at $g=15$ is more
loosely bound
than the $2$-bunch when
$g=20$. When $g=10$ the system is a completely unbound state. The case
$g=12$ is in the intermediate region, where the pair correlation still
has
a large peak at the nearest neighbor site, but does not show peaks at
positions which indicate the presence of a $2$-bunch. This value of
$g$ corresponds to a value of the potential such that the system is in
the
critical region of the phase diagram. In this region the $2$-bunch
ground
state
continuously unbinds into one with single bunches. 

We next consider the
case $u/g= 1.65$, where we study the ground state of 12 bosons on 120
sites. \fig{pairc2} shows that for large $g$ ($g=15$ and $g=7.5$) there is
clearly a $3$-bunch, which starts unbinding at $g=6$. At $g=5$ there is
a $2$-bunch. This starts to unbind around $g=4$ to a single bunch ground
state. It is
clear that
the cases $g=6$ and $g=4$ fall in the critical region, where the
unbinding of
a $3$ to $2$-bunch ground state and $2$ to $1$ bunch ground state takes
place, respectively. From \fig{pairc3}, for $g=2$, it is
evident that a $4$ bunch first unbinds to a $3$ bunch, then to
a $2$-bunch and finally a $1$-bunch. 

In \fig{pccomp}, we show points corresponding to
 Figs.~(\ref{pairc1})-(\ref{pairc3})
 on the phase
diagram derived for the extreme dilute limit (EDL).  Since these
points are for larger systems in the dilute limit, with multiple
bunches present and interacting, the precise phase boundary locations
for these systems are expected to differ from those for the EDL.  We
see that the behavior of the pairing correlation functions is
completely consistent with the conjectured phase diagrams.

\section{Equilibrium Crystal Shapes}

 In this section we will discuss the equilibrium shapes
of crystals that have step-step interactions with competing components.
The crystal shapes consists of both smooth and rough regions. The
latter
regions consists of step bunches separated from each other by
distances
depending on the local curvature of the surface. From the discussions
on the phases of vicinal surfaces, it is clear that the size of the
bunches depend on the temperature of the crystal. If the bunches have
size, say $n_b$, at low temperatures. the size progressively
decreases  as the temperature is increased. In the
temperature regimes where we can apply perturbative treatments to
compute the step energies, we will find the crystal shapes
for both cylindrical and 3-dimensional crystals.

\subsection{Cylindrical Crystals}

We start with the simple case of a cylindrical crystal where the
variations
in the shape occur in a 2-dimensional plane i.e, the crystal is
infinitely long
in the third direction. The crystal shape is obtained by  using the
projected free energy of a surface of orientation $\theta$ with respect
to
the reference surface using Wulff construction. If one
tunes the temperatures such that bunches of size $n_b$ become stable,
then, the crystal shape will consist of rough
regions that have widely separated bunches of size $n_b$.
The projected free energy for such a surface can be
expressed using \eqn{perte} as
\begin{equation}
f(s) = \gamma_0 + \eta_1|s| + \eta_3|s|^3 +
O(|s|^4),
\label{crfe}
\end{equation}
where the contribution $O(|s|^4)$ comes from the short-ranged attractive
interaction between the bunches. In the above
equation, we have introduced quantities $\eta_1 =
\left[\eta(T) + f(n_b,T)\right]/a_z$ and $\eta_3 =
\pi^2(k_BT)^2\lambda_b^2/(6n_b^4\Sigma a_z^3)$. The crystal shape $z(x)$
is given by
\begin{equation}
\lambda z(x) = \tilde{f}(\eta)|_{\eta=-\lambda x},
\end{equation}
where  $\tilde{f}(\eta) = min_s[f(s) - \eta s]$ is the Legendre
transform
of $f(s)$ to the conjugate field $\eta$, an $\lambda$ is parameter
fixing
the overall size of the crystal \cite{Noz}. The minimization leads to
the
result
\begin{eqnarray}
z(x) &=& 0, \,\,\,\,\, x < x_0, \nonumber \\
&=& -\frac{2\sqrt{\lambda}}{3\sqrt{3\eta_3}}[ x -x_0]^{3/2}, \,\,\,\,\
x>x_0,
\label{shape}
\end{eqnarray}
where $x_0 \equiv \eta_1/\lambda$. Notice that the crystal the rough
regions joins the flat region in a "continuous" with the well known
"3/2"
exponent. The above expression for crystal shapes holds in regions
were there is a bunch of a stable size, so that free energy
of the surface is obtained by perturbative methods. In the
critical region, where we do not have an analytical expression of
free energy, we will not be able to write down a simple expression
like
\eqn{shape}. However, from the general arguments presented in the
previous
section and the shape of the free energy shown in \fig{esplit}, we expect
the
shapes to evolve smoothly as the stable bunch size changes. Finally we note that at very high temperatures, where
the stable bunch size becomes one, the free energy reduces to
the standard free fermion energy and one recovers the
expression for the crystal shape obtained in Ref.~\cite{jrs}.

\subsection{3-Dimensional Crystals}

We now turn our attention from the cylindrical crystals to a
3-dimensional
crystal. In this case the crystal has flat facets that are joined to the
rough regions as shown in \fig{3dcrystal}. Once again we will focus on
temperatures such that the rough regions  have stable bunches of size $n_b$. The
orientation-dependent free energy $f({\bf p})$, where $\left(\partial z/\partial
x, \partial z/\partial y \right) = -|{\bf p}|(cos\phi,sin\phi)$, now takes the form
\be
f(|{\bf p}|) = \gamma_0 + \eta_1(\phi)|{\bf p}| + \eta_3(\phi)|{\bf
p}|^3 + O(|{\bf p}|^4),
\ee
where the coefficients $\eta_1$ and $\eta_3$ are functions
of the orientation $\phi$. Note that $\phi$ is the angle between the $y-$axis and
the average direction of steps at a given point on the surface. In the limit
$|{\bf p}|\rightarrow 0$, $\phi$ is the angle between the $y-$axis and the
direction of the facet edge. An important quantity that characterizes the three
dimensional crystal shape is the jump in gaussian curvature at the point where the
flat facet joins the rough region. The Gaussian curvature $K(x,y)$ of the crystal
at any general point ${\bf x} = (x,y)$ can be written as\cite{akutsu}
\be
K(x,y) = \left.\frac{\lambda^2(1+|{\bf p}|^2)^{-2}}
{det\left[\partial^2f(|{\bf
p}|)/\partial p_i\partial p_j\right]} \right|_{{\bf p}(-\lambda {\bf
x})},
\ee
from which the Gaussian curvature in at the facet edge $(|\bf p|
\rightarrow
0)$ can be cast in the form
\be
K(\phi) = {\lambda^2\over \left[6\eta_3(\phi)\left(\eta_1(\phi) +
\eta_1^{''}(\phi)\right)\right]}.
\ee
Using the fact that $\eta_1(\phi) +
\eta_1^{''}(\phi) = n_b\Sigma(\phi)/a_z$ and $\eta_3(\phi) =
\pi^2(k_BT)^2\lambda_b^2(\phi)/(6n_b^4\Sigma(\phi) a_z^3)$, the bunch stiffness, we find
that the jump in Gaussian curvature at the facet edge, $\Delta K$, is
 satisfies the relation
\be
\frac{k_BT\sqrt{\Delta K}}{\lambda} =
\frac{n_b^{3/2}a_z^2}{\pi\lambda_b(\phi)}.
\label{curv}
\ee
We see that the for ansisotropic crystals, where the elastic
interactions
between the steps depend on their orientation, the jump in Gaussian
curvature
depends on the angle $\phi$.
In the limit of high temperatures when $\lambda_b \rightarrow 1$ and $n_b = 1$,
we recover
the universal relation first derived by Akutsu {\it et al}\cite{akutsu}.
In the case when $U=0$, so that $n_b=1$, \eqn{curv} generalizes
the expressions that Saam\cite{Saam} obtained for certain specific
values of $g$,
namely, $g=0,-1/2$ and $
2$. Note that the jump in curvature in addition to being
non-universal depends not only on the strength
of the long range interaction, but also on the
strength of the short-range attraction through 
$n_b$. We note that in temperature window around which
the bunch size decreases, we expect the formula derived
for the $n_b$ bunch to evolve to the corresponding expression for
the the phase with reduced bunch size.
 However, in the absence of the exact form of free energy
in the critical region, we cannot study the details of this evolution.

\section{Comparison of our theory with continuum theories}

Recently the effect of competing short range and long range interactions
on
the finite temperature phase transitions on vicinal surfaces was
considered
by Lassig\cite{lassig} and Bhattacharjee\cite{bhat}. They studied
 a continuum version of the
model 
considered in this paper, where the fermions interact via a short range attractive
part
$-u$ of a range $a$
and an inverse square repulsive long range interaction of
strength $g$. 
In the continuum theory, all lengths are
measured in units of $a$, which can set to unity without loss of
generality. 
They analyzed the phase transition
using renormalization group(RG) techniques. For $N$ fermions on ring of
size $L$, such that $s \equiv N/L \rightarrow 0$, they find that
for a given $g$, when the short range attraction is
sufficiently attractive, the fermions
tend to phase separate. That is, for
$u>\frac{\sqrt{1+2g}}{2}$\cite{fac2}, the
fermions on the ring condense into a narrow region. With increasing $u$,
the
region's fermionic wave function becomes more and more localized, or in
other words
the steps try to form a {\em single} bunch of macroscopic size $N$. This
picture is
fundamentally different from our theory. In our picture, the steps do
not phase
separate to form a single macroscopic bunch. Instead, they form bunches of a finite size. The bunches then
fill the
space and interact with each other with renormalized interaction
parameters as
discussed in the text. With increasing temperature
the steps undergo a series of "peeling" transitions where the
bunch size progressively decreases by one.
As
mentioned earlier, the size of a bunch and peeling transitions are
lattice effects which are not present in the continuum model. In the
continuum model, steps (bosons) are treated as point objects. An
infinite number of these points can thus collapse to a space of
arbitrarily small size in order to maximize the short-range
attraction. Therefore, for the physical region we are concerned with,
namely when the range of the short-range attraction is comparable to
the ``size'' of the bosons, the continuum model is not valid. Due to
the underlying crystal structure, such a region may well be more
relevant for describing the experimental situation of surfaces.

For a fixed strength of the interaction, the continuum
theory
only captures the onset of the first bunching instability as
the temperature is decreased. In our
model, the first
bunching instability that is encountered on cooling from the high
temperature
phase is the point where 2-bunches first appear. For this transition, we
find
quantitative agreement with the continuum RG theory. In particular,
 the width of the critical region scales with density
like $\Delta T_c \sim s^{\sqrt{1+2g}}$\cite{lassig} for $g <3/2$, which is
precisely the relation that we obtain in \eqn{dg2}. For $g>3/2$,
the continuum theory fails to make any predictions because of
a singularity in the fermionic wave functions. However the natural
cutoff present in the lattice model, prevents such a problem in the
lattice model.

Our theory of step unbinding and the continuum theory also
give very different results for the crystal shapes, as illustrated in
\fig{crcomp}. Since the
steps phase separate in the continuum theory, there is
a slope discontinuity at the place where the facet of the crystal
joins the rough part at low temperatures. This discontinuity in slope
decreases continuously
and vanishes at the tricritical point. Above the tricritical
temperature, the rough part joins the facet with the $3/2$ exponent.
In our model the rough part joins the facet with the $3/2$ exponent
at all temperatures. As described earlier,
the sizes of the step bunches in the rough regions decrease with
increasing temperature.

\section{Comparison of theory with experiments on Silicon surfaces}

As noted in the introduction, one of 
the objectives of the this work is 
to provide an explanation for the experimentally observed 
exponents and for the physics of bunching phenomenona.
In principle, to accomplish this one would have to
carry out the calculations detailed in the text 
using 
 interaction potentials between steps derived from
electronic structure or other quantum mechanical calculations. Since
such calculations are not available at this time, we  
use the simple model with an attractive interaction restricted to
only near neighbor sites to explain the experimental observations.

\subsection{Experiments of Song et.al}
We first consider the experiments of Song et.al\cite{sm1,sm2,sm3},
where x-ray scattering studies were conducted on Si(113) miscut from
the
[113] direction towards [001].
 It was seen that depending on the slope of
miscut surface $s$, above a temperature $T_c(s)$, the surface is
uniformly stepped, while the surface consists of bunched steps below $T_c(s)$.
The number of steps in a bunch was seen to be about $22$. From the
measurements it was inferred that $s \sim |T_c(0)-T_c(s)|^{\beta}$,
with $\beta = 0.42 \pm .1$. Earlier work based on simple mean-field theory\cite{jrs},
predicted this exponent to be $\beta = 1.0$. 
Recent renormalization group calculations\cite{lassig,bhat} predict
that
the exponent depends on the ratio of the strengths of the attractive
and repulsive interaction strengths. In the region where these
calculations are valid, the exponent $\beta \ge 0.5$.
The exponent $\beta = 0.5$  corresponds to a special point
for 1-D systems interacting via the inverse square potential, a point beyond
which the renormalization group methods fail. The continuum theories
therefore require a particular value for $U/G$ to obtain agreement with
experiments.   Within our picture we identify the experimentally observed
transition as a 1-bunch to 22-bunch transition. While this transition
has not been explicitly studied in this paper, we argue that
the features of such a transition can be understood based on results
obtained in Sec.~IV. First note that like the 1-3 transition that
we observe for $u/g > 2.4$, a 1-22 transition can easily be expected
in a system with a short ranged potential with a more complex
structure. Even in a potential with next to near neighbor
interaction i.e. $n_s=2$ and with $U_2 > 0.25$,  at $T=0$,
one directly goes from a one bunch to a 3-bunch bypassing the
2-bunch. For potentials with more complicated features, it is
conceivable that one may go directly from a 1-bunch to a phase
with many steps in a bunch even at $T=0$. In these systems, one may easily
expect a 1-22 transition, just as we observed a 1-3 transition in
a simple model. Furthermore, the arguments presented in
Sec.~IV.  on the shift in transition temperature very well apply
to the 1-22 transition provided that the free energy curves $f(22,T)$ and
$f(1,T)$ intersect with a finite slope. If this is the case, we find
that the shift in transition temperature goes like $s^2$ from which
it follows that $\beta = 0.5$, in agreement with the
experimentally observed exponent.
Another key point is that when the bunch interactions are taken into
 account, we have argued that the transitions that are first order in
 the EDL become continuous. This means that the the exponent $\beta$
describes a curve of continuous 1-22 bunch transitions rather than a 
curve of first order transitions associated with a tricritical point.

  From an analysis of the x-ray scattering amplitudes Song et.al\cite{sm2} were
able to extract the temperature dependence of the stiffness
of the miscut surfaces.
Within our bunching picture the stiffness
of a bunched vicinal surface can be expressed as\cite{Noz}
$\tilde{\gamma}= \sqrt{\gamma_{||}\gamma_{\perp}}$, where
the stiffness along the direction of the steps can be written
as $\gamma_{||}= \Sigma_b/\theta$ while the stiffness in the
direction perpendicular to the steps is expressed in terms of the
surface energy of the miscut surface $\gamma(\theta)$ as
$\gamma_{\perp} = \gamma(\theta) + \gamma^{\prime
\prime}(\theta)$. Using \eqn{perte}, we can write the scaled surface
stiffness, $\hat{\gamma}$
in a stable $n_b$ bunch phase as
\be
\hat{\gamma} = \frac{a_z^2{\tilde{\gamma}}}{\pi k_BT} = \frac{\lambda_b}{n_b^{3/2}}.
\label{ss}
\ee
Using the value of scaled stiffness measured\cite{sm1} for the $2.1^0$
miscut
surface,
$ \hat{\gamma} = 1.5$ at
$(T-T_c)/T_c\approx 0.15$ (here $T_c$ is the temperature at
which the 22-bunch phase emerges)
 to represent the
stiffness of the $n_b=1$ phase as the critical region is approached and
using \eqn{ss}, we find $g(T_c) \approx 3/2$. Using this value of
$g$ and letting $n_b=22$, we find the scaled stiffness of the bunched
phase at $(T_c-T)/T_c=.15$ to be $1.3$. This is closer to the
experimentally observed value of $2.2$ than the value of $0.87$
predicted
by earlier mean-field studies\cite{jrs}. We note that there are
some
uncertainties
in the
the experimental results for stiffness in the 22-bunch phase and at
low temperatures, which might be the cause of the discrepancy between
our
results and the experimental observations.
In Ref.\cite{sm2}, the authors derive an exponent for the temperature
dependence of the stiffness by identifying a spinoidal temperature.
Within our picture the transition is continuous and a spinoidal
temperature does not exist. 

Finally, we note that Song {\it et al}\cite{sm3} have studied equilibrating bunches and
have sucessfully explained the limiting bunch size in terms of the Marchenko groove
picture discussed in Section III.A above. 

\subsection{Experiments of Sudoh et. al}

 We now turn to the recent experiments of Sudoh et.al\cite{sudoh}
 where the morphology of Si(113) surfaces, miscut towards a low
 symmetry
azimuth which is $33^0$ away from the $[\bar{1}10]$ direction to the
 $[\bar{3}\bar{3}2]$ direction
was studied using scanning tunneling microscopy. The authors 
compared surfaces quenched
 after annealing at various temperatures in the range
 600-1000$^0$C. Annealing
at temperatures above 720$^0$C resulted in single,
double, and triple-height steps,
 while
annealing at a temperatures in the range 690-720$^0$C the average
 terrace
spacing increased due to presence of double, triple, and quadruple-layer
steps. The size of the terraces saturate during annealing, indicating
 that
the surface had reached local thermal equilibrium. STM images at 710$^0$C
 show
a predominance of double steps. The authors also find that
the stiffness of the a step is proportional to its height, with an
 additional
stabilization of double height steps.

 Most of the features of the above observations can be understood
using the model considered in this paper, in which the attractive interaction is
restricted to the nearest neighbor site . We conjecture that the
temperature around 720$^0$C corresponds approximately to $u/g \approx 2.35$ and
$1/g
\approx 0.5$, where the stable phase is the 2-bunch phase. This point is marked
as "Su"
 in \fig{phasd}. Note that both the 1-2 line and 2-3 lines are very
 close
to this point, while the 3-4 line is also nearby. This means the free energies of the 1,2
and 3 bunches are very
 close
to each other with the free energy of the 2-bunch being the lowest,
explaining its
stability. Though the 2-bunch is stable in the EDL, the point Su is very
 likely
to be in the critical region, where the 1 and 3 bunch phases are
 stabilized
by bunch interactions. Further decreasing the temperature results in
 the
emergence of the 4-bunch phase, which can be understood by noting that
the system then moves closer to the 3-4 line resulting in the
stabilization of the 4-bunch due to bunch interactions. We also point out
our arguments that the stiffness of the bunched steps should be
 proportional
to the number of steps in the bunch  is also borne out in the experiments. 
Also, note that
this vicinal surface oriented in a direction different from that of
Song et.al. As a result we expect different strengths of attractive
and repulsive step interactions leading to different bunch sizes in the two
experiments.

Step bunches with $n_b$ in the range 1-4 have also been seen on miscut Si(113) by other
workers, specifically van Dijken {\it et al}\cite{vand} and Zhu {\it et al}\cite{zhu}.
Equilibrium does not appear to have been achieved in these cases.
 
\section{Discussion and Future Directions}
In this paper we have explored the phase transitions arising in
vicinal surfaces with competing long-range and short range
interactions.
We showed that depending on the strength of the interactions and
the
temperature, the steps on the surface rearranged themselves into
bunches. Using a tractable model for
step interactions, we showed that the bunch size increased with
decreasing temperature through a series of bunching transitions.
At $T=0$, we showed that the bunch phases predicted by our model
correspond to well known periodic groove structure first predicted by
Marchenko. Our theory can then be considered as an extension of
the Marchenko theory to finite temperatures, where changes in
the size and
number of steps in a groove are brought about by the bunching transitions.
The implications to experiments and the physics of 
crystal shapes was pointed out. We found that the  exponent
relating the dependence of the transition temperature
on the miscut angle, $\beta$, expected from our theory to be consistent with 
the experimentally measured exponent
and that the observed features of step stiffness was also in agreement with
our theory. We now point out some directions for future theoretical and
experimental research that
can provide more insights  on equilibrium bunching on vicinal
surfaces.

 On the theoretical/computational side, it is very important to have
 an
accurate determination of the step interaction potential. This can be
 done through quantum calculations or atomistic simulations using
 reliable
empirical inter-atomic potentials. Once the step interactions are
 known, the methods developed in this paper can be used to study the
step bunching transitions. 
On the experimental side more data on surfaces with competing
interactions
can shed light on the role of attractive interactions on equilibrium 
bunching properties. 
In particular for vicinal surfaces in a
bunched phase, the crystal shape will exhibit a Pokrovsky-Talapov
transition as the facet is
approached. This is also true within the Marchenko picture. An
experimental check of this feature would
be valuable. It would also be useful to determine the exponent $\beta$
for a number of different surfaces that have attractive interactions
 between steps. In contrast to the recent RG calculations that require
a particular ratio of the attractive to repulsive interaction
 strengths to obtain $\beta = 0.5$,
our analysis shows that this value of the exponent is robust, applying
 to
all the bunching transitions that are first order in the EDL.   

A detailed study of the step phases as a function of orientation of the silicon
surfaces away from the [113] direction would be most useful in order to elucidate
the connection between the experiments of Sudoh {\it et al} and those of Song {\it
et al}. Note that miscuts such as those in the former experiments introduce kinks
in steps on the Si(113) surface. For small kink densities, the model parameters
$G$, $U$, and $\eta$ should have corrections proportional to the kink density. The
phase boundaries in \fig{phasd} will then shift linearly with kink density,
providing a means of tuning the model parameters. For cases such as that marked by
Su in \fig{phasd}, small changes in the kink density would lead to pronounced
effects due to the close proximity of phase boundaries.

The quantum n-bunch, or n-mer, phases discovered in this work are highly
interesting in their own
right, independent of their realization in crystal surface physics.
While we expect these phases to be
Luttinger liquids well away from transitions from one bunch phase to
another, the picture may change
in the crossover region where there is a mixing of the long-wavelength
modes associated with
phonon-like oscillations of the bunches and the modes internal to the
individual bunches. 
In effect, new many-body phases may emerge in the critical
region.
A detailed study of the crossover
region will shed light on some these new physical phenomenon. 

\section{Acknowledgments}

VBS would like to thank  Sanjay Khare and Avraham Schiller
for stimulating discussions. The Quantum Monte Carlo Simulations were
performed on the IBM SP2 at the Ohio Supercomputer Center.

\appendix

\section{ Free energy of an inclined single step in the TSK
model}
In this appendix we derive the free energy of a single step placed on a
rectangular lattice with lattice constants $(a_x,a_y)$. The step runs on the
average at an angle
$\alpha$ with respect to the y-axis (The straight step has $\alpha = 0$.). The
slope of the step with respect to the y-axis is then $\hat{s}=\tan\alpha$. If
the free energy per unit length of the step as a function of $\hat{s}$ is
$\hat{f}(\hat{s})$, then in the expansion for small $\hat{s}$
\begin{equation}
\hat{f}(\hat{s})=\hat{f}_0+\frac{1}{2}\Sigma\hat{s}^2+\cdots,
\label{a1}
\end{equation}
$\Sigma$ is the step stiffness. To compute $\hat{f}(\hat{s})$, it is easier
to first compute the Legendre-transformed free energy 

\begin{equation}
\hat{f}(\hat{\eta})=\hat{f}(\hat{s})-\hat{\eta}\hat{s},
\label{a2}
\end{equation}
where $\hat{\eta}$ is a field coupling to the slope $\hat{s}$\cite{Noz}.
We assume that in going from one row to the next, the step can follow only
three paths. It can go straight to the next row up with energy $\epsilon_0$,
to the left one column and then up one row with total energy
$\epsilon_0+\Gamma-\hat{\eta}$, or to the right one column and then up one
row with  total energy $\epsilon_0+\Gamma+\hat{\eta}$. The partition function
for the step is readily found, yielding the free energy

\begin{equation}
a_y\hat{f}(\hat{\eta})=\epsilon_0-k_BT\ln[1+e^{-\beta(\Gamma-\eta)}+
e^{-\beta(\Gamma+\eta)}]
\label{a3}
\end{equation}
Combining Eqs.(\ref{a1}-\ref{a3}) yields (noting that $e^{-\beta\Gamma}\ll 1$
for this model to apply), we obtain \eqn{gamma} of the text.

\section{Quantum Monte Carlo Methods}

\noindent {\bf Variational Monte Carlo method:}
This is the simplest way of performing a quantum Monte Carlo simulation. Here,
our
aim is to have a trial wave function that has a set of parameters that
can
be varied. One then optimizes these parameters so as to obtain the
lowest
possible energy. By means of a Monte Carlo simulation, one can
compute the energy of a wave-function as a function of the parameters.

We start  with writing the trial state $|\psi_T>$  in terms of the
configurations of the
one dimensional lattice system
\begin{equation}
|\psi_T> = \sum_R |R><R|\psi_T>,
\end{equation}
where the coefficients $<R|\psi_T>$ depend on the set of variational
parameters ${p_i}$. Here, $|R>$ is a state in configuration space. For
example, for N bosons on L sites a configuration could be a state with
bosons
on lattice sites $\left[i_1,i_2 ..i_N\right]$ such that $1 \le i_k \le
L$ for $1 \le k \le N$. 
 The expectation value of the energy in the state $|\psi_T>$ can
be
expressed as
\begin{eqnarray}
E_T &=& \frac{<\psi_T|{\cal H}|\psi_T>}{<\psi_T|\psi_T>} \nonumber \\
&=& \frac{\sum_R E_L(R)<\psi_T|R><R|\psi_T>}{\sum_R<\psi_T|R><R|\psi_T>},
\label{Evmc}
\end{eqnarray}
where the local energy in configuration $R$ is defined as
\begin{equation}
E_L(R) = \frac{<\psi_T|{\cal H}|R>}{<\psi_T|R>}.
\label{etr}
\end{equation}
The sum in Eq.~(\ref{Evmc}), in general impossible to evaluate
analytically, can be evaluated by Monte Carlo.
One considers the following expression:
\begin{equation}
E_T = \sum_R E_L(R)p(R),
\end{equation}
where one interprets
\begin{equation}
p(R) = \frac{<\psi_T|R><R|\psi_T>}{\sum_R<\psi_T|R><R|\psi_T>}
\end{equation}
 as a probability density function.
If one then samples $M$ configurations distributed
according
to $p(R)$ then the exact expression for $E_T$ can be approximated by
\begin{equation}
E_{VMC} = \frac{\sum_R^{\prime}E_L(R)}{M},
\label{EVmcEst}
\end{equation}
where the prime is used to represent sum over the configurations. In the
limit of large $M$, $E_{VMC}$ will converge to
$E_T$. The average $E_{VMC}$, of course, is subject to statistical
fluctuations, and one can easily evaluate the statistical error by
$\delta E = \sigma_E/\sqrt{M}$. The variance of the local energy,
$\sigma_E$, is defined as
\begin{equation}
\sigma_E = \sqrt{<E^2>-<E>^2} ,
\end{equation}
where $<E> = E_{VMC}$ and $<E^2> = \frac{\sum_R^{\prime}E_L^2(R)}{M}$.
It is known\cite{cyrus} that $\sigma_E$, which has a lower bound 
of $0$, is a better quantity to 
optimize than $E_{VMC}$. We use both quantities as indicators 
in our search of variational parameters.

For each set of variational parameters,
we use the Metropolis algorithm to generate a set of $M$
configurations
distributed according to $p(R)$. We start with a configuration of $M$
walkers each of which has $N$ bosons randomly distributed on a lattice
of
one dimensional ring of size $L$. Each walker then undergoes a
"stochastic
walk" according to the following procedure: (1) for each boson we pick a
near
neighbor site randomly and (2) if the site is occupied the boson is not
moved,
but if the site is empty the boson is moved with probability $q$ given
by
\begin{equation}
q = min\left[ 1, \frac{\psi_T^2|_{new}}{\psi_T^2|_{old}}\right],
\end{equation}
where $\psi_T^2|_{new(old)}$ is 
$\langle \psi_T|R\rangle\,\langle R|\psi_T\rangle]$ 
evaluated at $R_{new}$ and $R_{old}$.
This process is repeated  sufficient number of times so that one ends up
with a set of walkers distributed according to $p(R)$.

\noindent {\bf Green's Function Monte Carlo method:}
This is in principle an exact way to calculate the ground state
properties
of Bose systems. Its required computer time scales algebraically 
(as opposed to exponentially in exact diagonalization) with
system size, thus allowing exact calculations for large systems.
Here, one combines random sampling with a simple method
to project out the ground state from an arbitrary initial state. 

We start with the operator
\begin{equation}
{\cal F} \equiv C - {\cal H},
\end{equation}
where
the constant $C$, whose choice we discuss below, 
is to ensure that all matrix elements 
$\langle R^\prime |{\cal F} | R \rangle$ are non-negative.
The basic premise of GFMC is that repeated application
of ${\cal F}$ on an essentially arbitrary initial state 
will project out the ground state. That is, if an
initial state $|\psi^{(0)}>$ has any overlap with the ground
state $|\psi_0\rangle$ of ${\cal H}$, the process
\begin{equation}
|\psi^{(n)}> = {\cal F}^n|\psi^{(0)}>
\label{GFMCiter}
\end{equation}
will lead to $|\psi_0\rangle$ at large $n$. GFMC is a method 
to realize the above process by a Monte Carlo random walk.
 
In order to improve efficiency of the random walk process,
one more mathematical manipulation of Eq.~(\ref{GFMCiter})
is necessary. This is done by introducing 
an operator $\tilde{\cal F}$ whose matrix elements are 
related to those of ${\cal F}$ by a similarity transformation:
\begin{equation}
\tilde{\cal F}(R^\prime,R) \equiv \langle \psi_T|R^\prime\rangle
\,\langle R^\prime |{\cal F}| R\rangle/\langle \psi_T|R\rangle.
\end{equation}
The stochastic realization of Eq.~(\ref{GFMCiter}) is actually
with $\tilde{\cal F}$ instead of ${\cal F}$. While mathematically
equivalent, the use of $\tilde{\cal F}$ can significantly reduce the fluctuation 
of the Monte Carlo process if $|\psi_T\rangle$ is a reasonable
approximation of $|\psi_0\rangle$. This is known as importance
sampling\cite{GFMC}.

The program
is then
to start with a set of walkers distributed according to
$<\psi_T|R><R|\psi_T>$. 
These walkers undergo random walks
in $R$-space. For each walker, denoted by $R$, taking a step
in the random walk means randomly selecting and moving to 
a new position $R^\prime$ with
probability 
$\tilde{\cal F}(R^\prime,R)/\sum_{R^\prime} \tilde{\cal F}(R^\prime,R)$.
Note that,
due to
the structure of ${\cal H}$,  
there are only ${\cal O}(N)$ possible $R^\prime$'s and the corresponding
probabilities can be easily computed. Because the overall normalization
$\sum_{R^\prime} \tilde{\cal F}(R^\prime,R)$ is not a constant,
each walker carries an overall weight which fluctuates, or a branching
scheme is introduced to allow the total number of walkers to fluctuate, 
or both.

The ground state energy is given exactly
by the so-called {\em mixed estimate}
\begin{eqnarray}
E_0 &=& \frac{<\psi_T|{\cal H}|\psi_0>}{<\psi_T|\psi_0>} \nonumber \\
&=& \frac{\sum_R E_L(R)<\psi_T|R><R|\psi_0>}{\sum_R<\psi_T|R><R|\psi_0>},
\label{gfe}
\end{eqnarray}
where $E_L(R)$ is defined in Eq.~(\ref{etr}). 
After a sufficient number of steps, the walkers are distributed according to
$<\psi_T|R><R|\psi_0>$. 
$E_0$ can therefore be computed from such walkers as the (weighted)
average of $E_L(R)$ with respect to walker positions $R$, similar to
Eq.~(\ref{EVmcEst}). We denote this Monte Carlo estimate of $E_0$ by
$E_{GFMC}$.

Expectation values of operators other than ${\cal H}$ can also be 
computed from the mixed estimate. However, if the operator does not 
commute with ${\cal H}$, this estimate is {\em not\/} exact. 
As we mention in Section \ref{QMCresults}, this is
an important distinction which requires careful analysis of the 
results. There exist ways to improve upon the mixed estimate and 
to possibly extract exact estimates of expectation 
values\cite{GFMC,bilinear}, 
but we will not discuss them here.

Now we describe our choice of $C$. Since we are dealing
with
a finite number of hard core bosons on a lattice of finite size, the
eigenstates of ${\cal
H}$ are bounded both from above and below. If $E_{max}$ is the highest
eigenvalue of ${\cal H}$, we choose $C$ such that $C > E_{max}$. This
is easily achieved for the problem at hand. On the lattice we put all
bosons next to one another and compute the potential energy arising from
the long range interactions alone. For $N$ bosons, we choose $C$ as
\begin{equation}
C = 2N + g\sum_{i=1}^{N-1}\frac{(N-i)}{i^2}.
\end{equation}
This clearly satisfies $C > E_{max}$, since hopping and the short range
attraction can only lower this energy.

\newpage
\center{{\bf FIGURE CAPTIONS}}

\begin{figure}
\caption{Geometry for the TSK model. The steps between terraces are indicated
by bonds on the lattice.They run parallel to the y-axis on an average.}
\label{steps}
\end{figure}

\begin{figure}
\caption{T=0 phase diagram of the vicinal surface plotted as a
function
of the ratio of the near neighbor attractive interaction $U$ and
repulsive inverse square interaction $G$. The dots separate regions
with stable bunch sizes that differ by one (The bunch sizes are given
indicated in the figure).}
\label{t=0pd}
\end{figure}

\begin{figure}
\caption{Plot of the difference in energy per boson in a $n_b$ bunch
and the energy of a 1-bunch. At the points indicated by dots, the
bunch
size corresponding to the minimum of the energy difference increases
by one.} 
\label{ened}
\end{figure}

\begin{figure}
\caption{Periodic Groove structure proposed by Marchenko. The figure
shows 
the repeating groove structure (length $L_g$), the flat facet (length
$L_1$) and the stepped facet (length $L_2$). The case $n_b=3$ is shown here for
graphical simplicity. Marchenko's calculation applies only for $n_b$ large.}
\label{marchenko}
\end{figure}

\begin{figure}
\caption{ Energy/step in bunches of size 2(circle), 3(square) and 4(triangle)
for
$u/g = 1.65$, plotted as a function of $g$. The energies are obtained
by
exact diagolization of the quantum bunch problem on 12 sites in the
case of 3 and 4 bunches and 200 sites for the 2 bunch.}
\label{eneps}
\end{figure}

\begin{figure}
\caption{ Phase diagram in the extreme dilute limit plotted in the
$u/g$-$1/g$ space. The lines separate regions with stable bunch sizes
that differ by one. The stable bunch size in each region is indicated
in the figure. The 1-2 line
(dotted line) lies above the 1-3 line (short-dashed line) for $u/g < 2.4$,
while
lies beneath it for $u/g <2.4$. For $u/g > 2.4$ the 1-3 line lies
in between the 1-2 and 2-3 (solid line) lines. This implies that for 
$u/g > 2.4$, the 1-bunch phase directly undergoes a transition to the
3-bunch phase bypassing the 2-bunch phase. The long-dashed line is the
3-4 line.}
\label{phasd}
\end{figure}

\begin{figure}
\caption{Comparison of the exact diagonalization calculations and GFMC
simulations.
The circles indicate the energy per step in a 3-bunch
confined
to 12-sites and for a 2-bunch on 200 sites as found from exact diagonalization, while the
triangles and squares represent the results for the 2-bunches and 3-bunches on 30 sites
and 100 sites, respectively, as found using GFMC. The
error bars on the GFMC energies are shown in the figure.  Lines are
drawn through the points to  guide the eye.}
\label{exadc}
\end{figure}

\begin{figure}
\caption{  Plot of the 1-3 line
and 3-4 line for $u/g < 8$.}
\label{phasd2}
\end{figure}

\begin{figure}
\caption{Schematic plot of the energy per step (solid curves) when
the bunch-bunch interactions are taken into consideration. The
dotted lines indicate the energy per step in the EDL. In Region
1,
the $n_b$-bunch has lower energy while in Region 3 the $n_b-1$ bunch
is lower in energy. In the critical region, the energy per step
smoothly passes from $f(n_b,T)$ to $f(n_b-1,T)$. The width of
the critical region indicated in the figure is estimated in the text.}
\label{esplit}
\end{figure}

\begin{figure}
\caption{Schematic depiction of the two body correlation $f(r)$ used
in the trial wave-function. The function $f(r)$ takes
on a value $f_1$ at $r=1$ has a minimum value of $d$ at $r=r_0$ and
behaves like $1-const/r^{\alpha}$ for large $r$.}
\label{2bcor}
\end{figure}

\begin{figure}
\caption{The boson pair correlation function $g(x)$  for 10 bosons
on 100 sites with $u/g = 1.25$ for values of $g$ indicated in the
figure. The points are also marked in \fig{pccomp} for comparison.
By counting the number of peaks in $g(x)$ and making use of the
periodic boundary condition the number of bosons in a bunch
can be established. For example, in the case $g=20$, there are
2 peaks at $x=20$ and $x=40$ in addition to a large value of $g(x)$
on the near neighbor site. Invoking the periodic boundary condition,
one sees that if there are $2$ bosons in a bunch, the four peaks
at $x=\pm20$, $x=\pm40$ account for 8 bosons, which along with
the boson on the near neighbor site and the boson at the origin
add up to 10 bosons. 
For $g=20$ and $g=15$ we have a well defined 2-bunch phase, while
there is a well defined 1-bunch phase for $g=10$. The figure shows
that $g=12$ is in the critical region. We illustrate the statistical
error bars
only
for the case $g=20$, since in all the other cases they have
similar magnitudes.}
\label{pairc1}
\end{figure}

\begin{figure}
\caption{Boson pair correlation function (\eqn{corf}) for 12 bosons
on 120 sites with $u/g = 1.65$ for values of $g$ indicated in the
figure. The points are also marked in \fig{pccomp} for comparison.
Counting the number of peaks as we did in \fig{pairc1} gives
the number of bosons in a bunch. To account for the 3-bunches, note
that
when
$g=15$ there are 2 peaks at $x=30$ and $x=60$ in addition to 
large values of $g(x)$ on the near-neighbor and next to near-neighbor
sites. Invoking the periodic boundary condition, we see that the peaks at
$x=\pm30$ account for 6 bosons while the peak at $x=60$ accounts
for 3 bosons which along with the boson at the origin and the ones on
two adjacent sites account for all 12 bosons.
For $g=15$ and $g=7.5$ we have a well defined 2-bunch phase, while
there is a well defined 2-bunch phase for $g=5$ and 1-bunch phases
for $g=4$ and $g=3$. The figure shows
that $g=6$ is in the critical region. The statistical 
error bars for all values
of $g$ have about the same magnitude as the case $g=15$ where
they have been explicitly displayed.}  
\label{pairc2}
\end{figure}

\begin{figure}
\caption{Boson pair correlation function (\eqn{corf}) for 12 bosons
on 120 sites with $u/g = 2.0$ for values of $g$ indicated in the
figure. The points are also marked in \fig{pccomp}  for comparison.
 By counting the number of peaks in $g(x)$ as we did
in the previous figures, we see that for $g=7.0$, $g=3.5$, $g=2.85$ and $g=2.25$  we have  well defined 4,3,2 and 1-bunch phases respectively, while
the points $g=4$ and $g=2.96$ are in the critical region. The error
bars
have only been shown for the case $g=7$, since in all the other
cases they have similar magnitudes.}  
\label{pairc3}
\end{figure}

\begin{figure}
\caption{Comparison of the phase diagrams obtained in the EDL using
exact diagonalization and GFMC simulations. For $u/g=1.25, 1.65,2.0$
the pair correlation functions for the marked values
of $g$ are given in Fig.~(\ref{pairc1})-(\ref{pairc3}). For each value
of $g$ we also indicate the bunch size (1b,2b,3b and
4b correspond to 1,2,3 and 4 bunches respectively,
while cr refers to the critical region) obtained from GFMC simulations.
Note that the GFMC results and the exact diagonalization results
do not agree exactly since in the dilute limit the
bunch interactions can alter the nature of the phases especially
close to the phase boundaries.}
\label{pccomp}
\end{figure}

\begin{figure}
\caption{Top view of a 3-d crystal. The flat facets and the rough
regions
are shown in the figure. The
steps in the rough region run parallel to the direction inclined
to the vertical by an angle $\phi$ as shown, for a point at the facet edge, in the
figure. The z-axis points outside the plane of the paper as indicated in the
figure.}
\label{3dcrystal}
\end{figure}

\begin{figure}
\caption{Comparison of the temperature dependence of crystal shapes
in lattice and continuum theories. In the lattice theory the
rough region always joins the facet continuously with the standard
``3/2'' exponent. In this theory, the rough regions consists of
steps in bunches, whose sizes decrease with increasing
temperature. Within
the continuum theory, the rough regions joins the facet with a
discontinuous slope. This discontinuity decreases with increasing
temperature
and vanishes at the tricritical point.}
\label{crcomp}
\end{figure}

\newpage
\center{{\bf TABLES}}
\bigskip
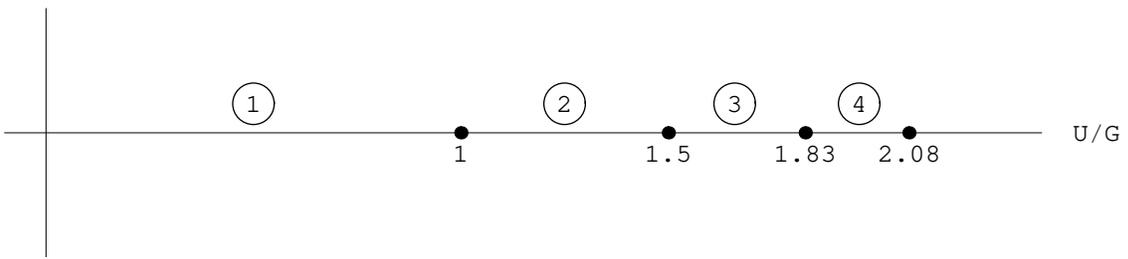
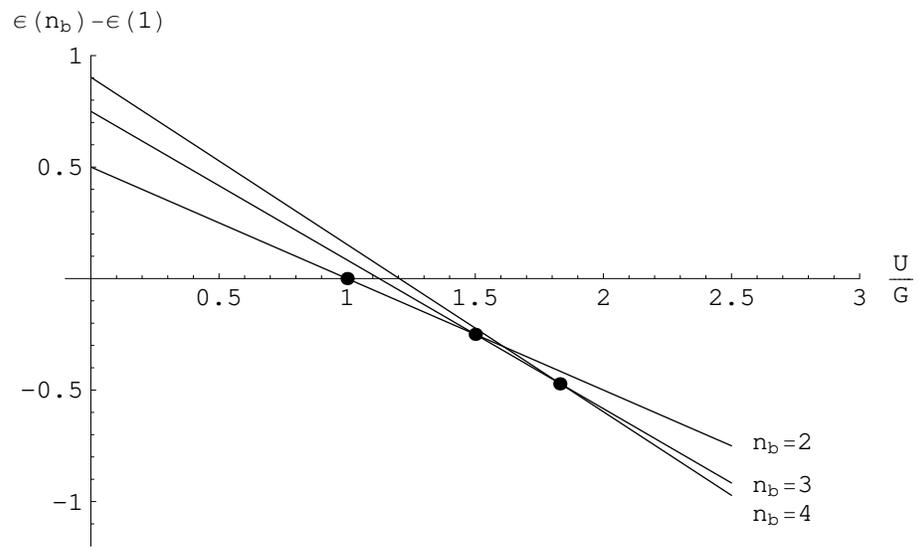
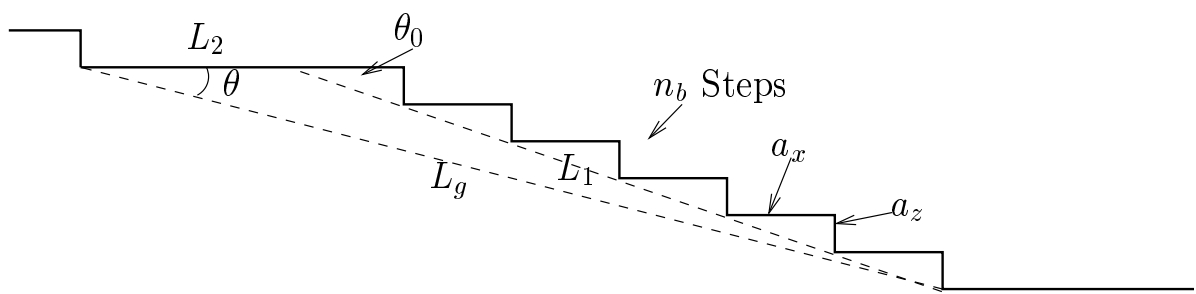
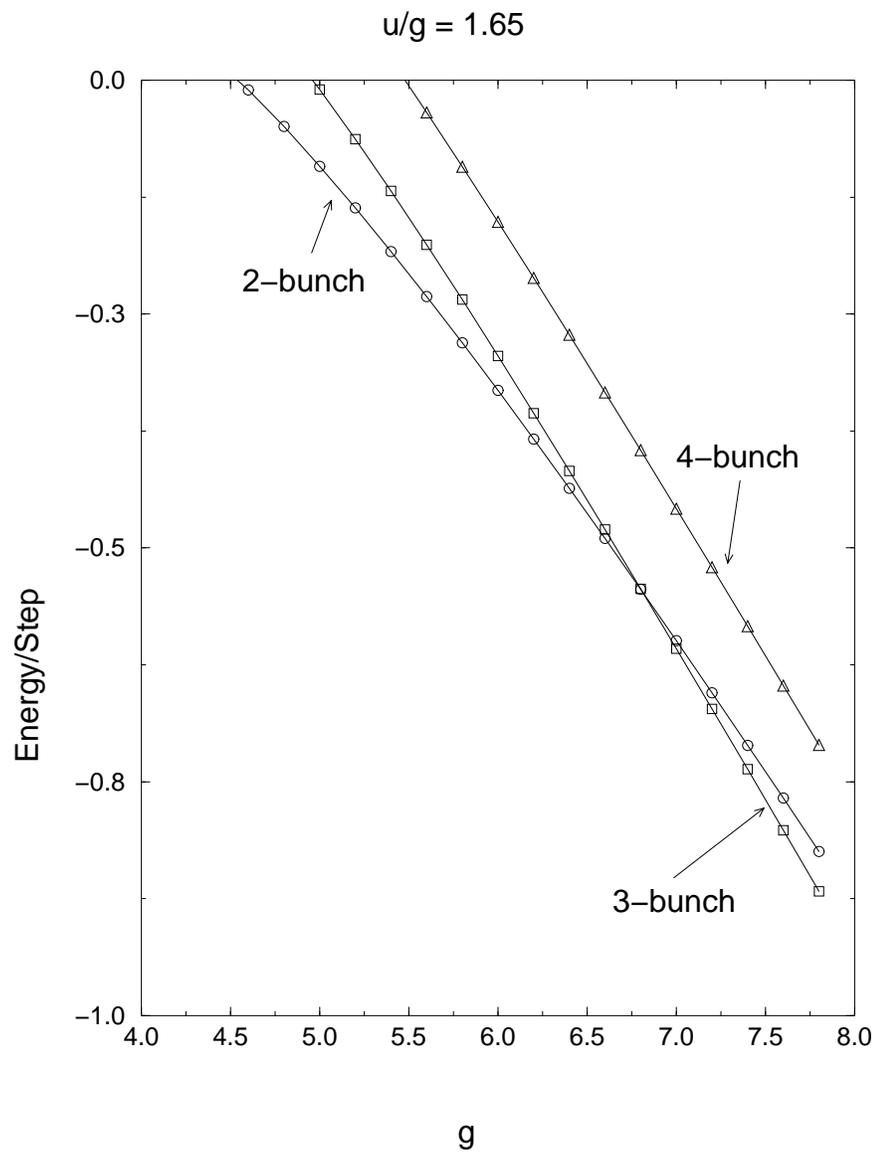
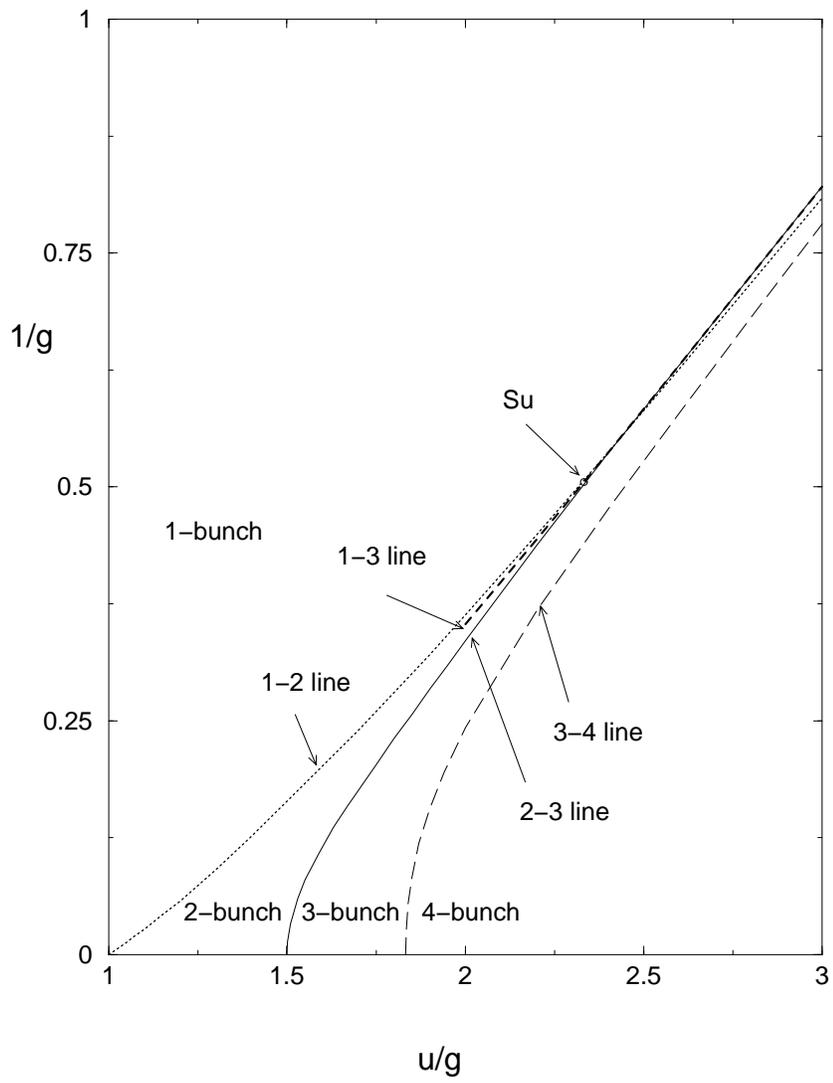
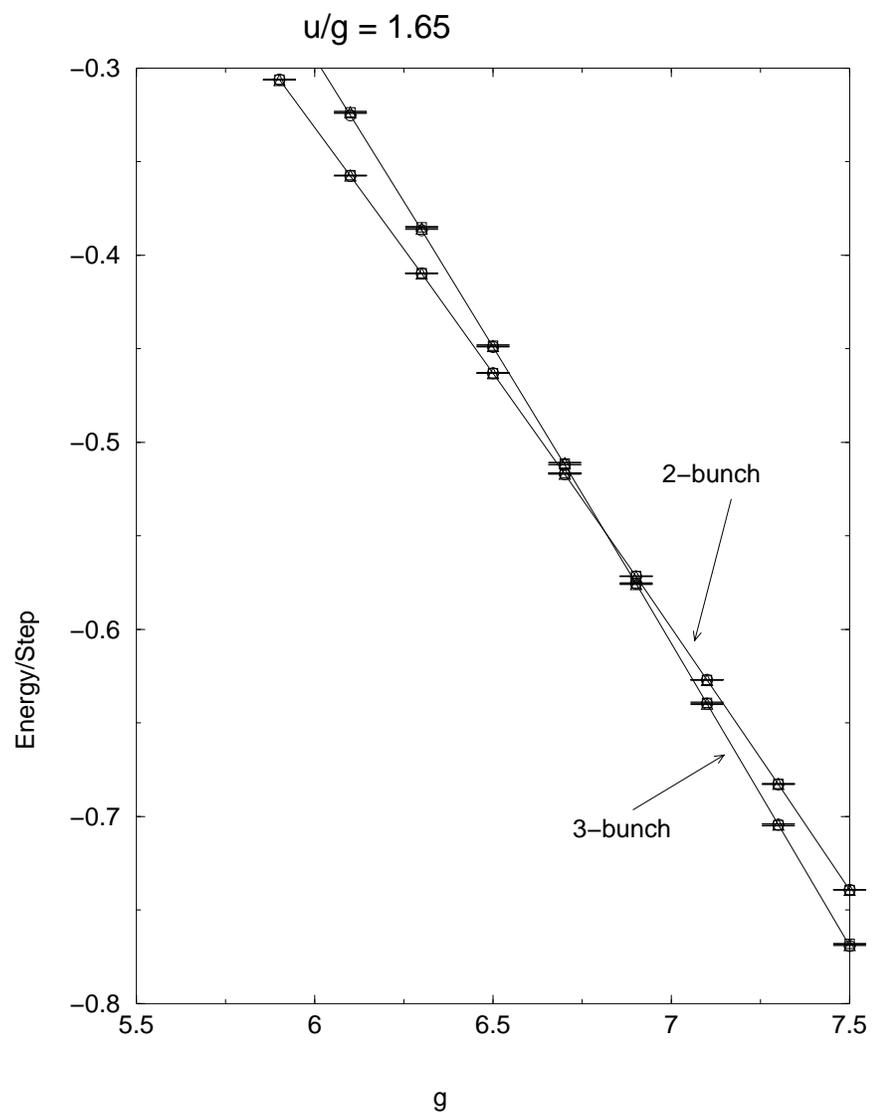
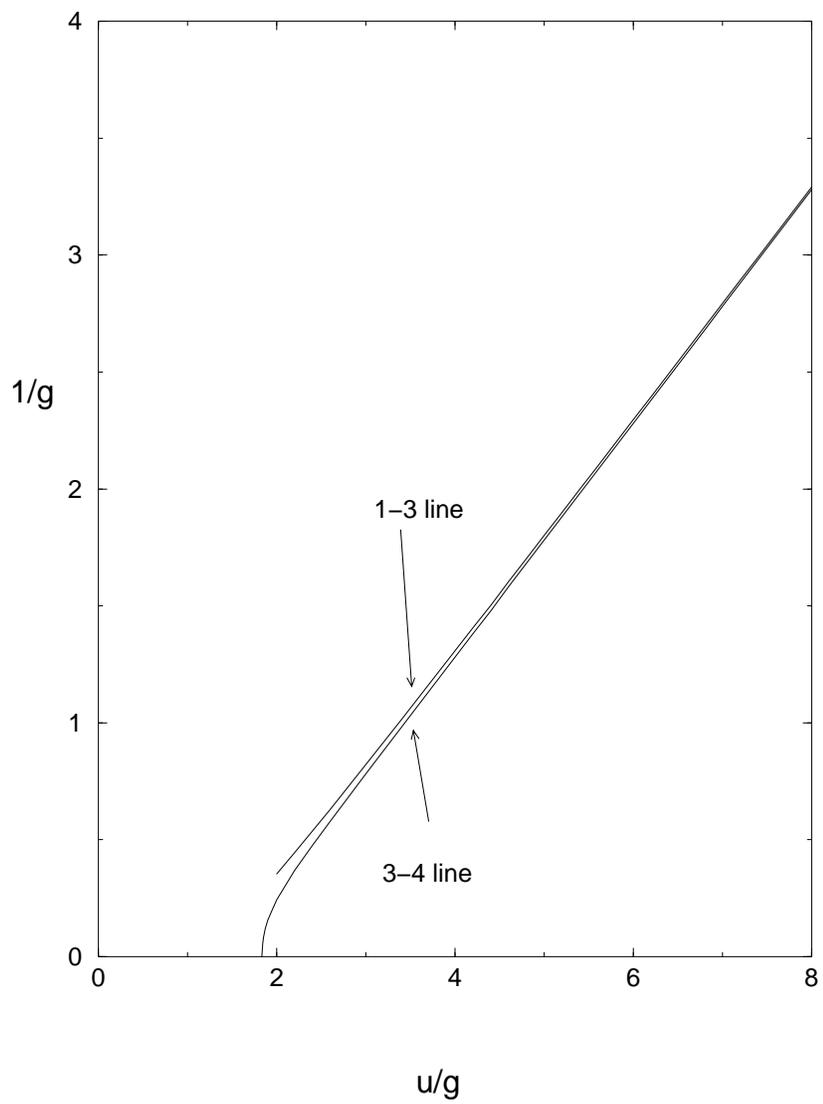
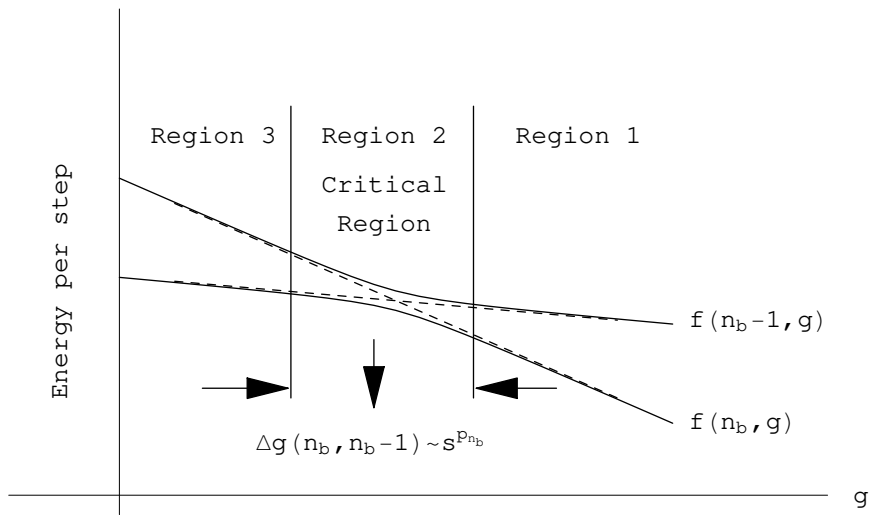
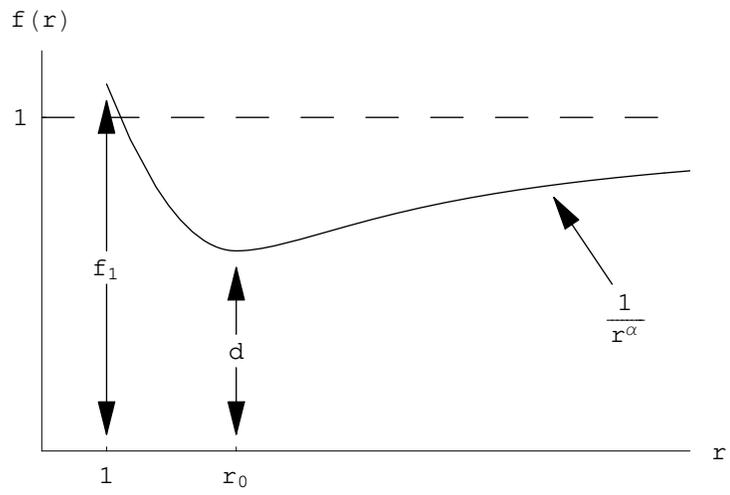
\widetext
\begin{table}
\caption{Comparison of the exact diagonalization and GFMC results for
3 bosons confined on 12 sites with $u/g=1.65$. $E_{exad}$ denotes the
energy per boson obtained by exact diagonalization, $E_{var}$ is the
variational energy per boson on 12 sites and $E^{12}_{GFMC}$
represents
the GFMC result. The number bracket, corresponding to
the GFMC energies, indicates the error on the last decimal
place.
In all the cases a variational wavefunction with $r_0=3$, $f_1=1.14$,
$d=0.5$ and $\alpha=0.6$ was chosen. We also give the GFMC energies
 ,$E^{30}_{GFMC}$ and$E^{100}_{GFMC}$  ,
for 3 bosons  occupying 30 and 100 sites respectively. }
\begin{tabular}{cccccc}
$g$ & $E_{exad}$ & $E_{VMC}$ & $E^{12}_{GFMC}$ &
$E^{30}_{GFMC}$&$E^{100}_{GFMC}$\\
\tableline
6.5 &  -0.44902 & -0.34(3) & -0.4489(2) & -0.4490(4) & -0.4492(7)\\
6.9 &  -0.57562 & -0.47(3) & -0.5756(1) &  -0.5756(4)&  -0.5753(6)\\
7.3 & -0.70425 & -0.63(4) & -0.7042(1) &  -0.7043(4)&  -0.7046(6)\\
\end{tabular}
\label{table1}
\end{table}

\begin{table}
\caption{Energies for $u/g = 1.25$ and values of
$g$ shown in the table, computed for 10 bosons 
on 100 sites. The variational energy is denoted
by $E_{VAR}$, the GFMC energy by $E_{GFMC}$ and $E_{approx}$ is
the approximate energy computed using \eqn{app}. The variational
wave function parametrized by $r_0$, 
$f_1$, $d$ and $\alpha$ (see \eqn{varwf}) is also tabulated.}
\begin{tabular}{cccccccc}
$g$ & $r_0$ & $f_1$ & $d$ & $\alpha$ & $E_{VMC}$ & $E_{GFMC}$ &
$E_{approx}$\\
\tableline
10.0 & 2.0 & 1.05 & 0.58 & 1.5 & 3.63(4) & 2.312(1) & 3.071\\
12.0 & 2.0 & 1.06 & 0.58 & 1.5 & 3.76(5) & 2.221(1) & 3.553\\
15.0 & 2.0 & 1.06 & 0.30 &  0.5 & 2.37(6) & -0.479(1) & -0.424\\
20.0 & 2.0 & 1.14 & 0.18 & 0.5 & -0.10(1) & -6.701(2) & -6.223\\
\end{tabular}
\label{table2}
\end{table}

\begin{table}
\caption{Energies for $u/g = 1.65$ and values of
$g$ shown in the table, computed for 12 bosons 
on 120 sites. The symbols have same meanings as in Table. II}
\begin{tabular}{cccccccc}
$g$ & $r_0$ & $f_1$ & $d$ & $\alpha$ & $E_{VMC}$ & $E_{GFMC}$ &
$E_{approx}$\\
\tableline
3.0 & 4.0 & 1.02 & 0.7 & 1.0 & 1.99(2) & 1.139(1) & 1.311\\
4.0 & 4.0 & 1.02 & 0.3 & 0.4 & 1.85(4) & 1.032(1) & 1.579\\
5.0 & 4.0 & 1.02 & 0.4 & 0.4 & 0.40(3) & -0.620(1) & -0.486\\
6.0 & 4.0 & 1.14 & 0.1 & 0.1 & -1.68(9) & -3.340(3) & -3.363\\
7.5 & 4.0 & 1.14 & 0.1 & 0.1 & -6.83(7) & -8.870(7) & -8.742\\
15.0 & 4.0 & 1.16 & 0.1 & 0.1 & -33.77(3) & -39.075(3) &-38.814\\
\end{tabular}
\label{table3}
\end{table}

\widetext
\begin{table}
\caption{Energies for $u/g = 2.0$ and values of
$g$ shown in the table, computed for 12 bosons 
on 120 sites. The symbols have same meanings as in Table. II}
\begin{tabular}{cccccccc}
$g$ & $r_0$ & $f_1$ & $d$ & $\alpha$ & $E_{VMC}$ & $E_{GFMC}$ &
$E_{approx}$\\
\tableline
2.25 & 3.0 & 1.04 & 0.7 & 0.7 & 1.11(1) & 0.807(1) & 1.104\\
2.85 & 3.0 & 1.06 & 0.5 & 0.1 & 0.89(1) & 0.051(5) & 0.119\\
2.96 & 4.0 & 1.02 & 0.1 & 0.1 & 0.34(2) & -0.301(1) & -0.209\\
3.5 & 4.0 & 1.02 & 0.1 & 0.1 & -2.31(3) & -3.181(1) & -3.160\\
4.0 & 4.0 & 1.06 & 0.1 & 0.1 & -4.61(1) & -6.165(5) & -6.144\\
7.0 & 4.0 & 1.06 & 0.1 & 0.1 & -24.21(3) & -26.779(1) & -26.665\\
\end{tabular}
\label{table4}
\end{table}


\begin{references}

\bibitem{sm1} S. Song and S. G. J. Mochrie, {\it Phys. Rev.
Lett.} {\bf 73}, 995 (1994).

\bibitem{sm2} S. Song and S. G. J. Mochrie,
{\it Phys. Rev. B} {\bf 51}, 10068 (1995).


\bibitem{sm3} S. Song {\it et al},
{\it Surf. Sci.} {\bf 372}, 37 (1997).

\bibitem{sudoh} K. Sudoh {\it et al}, {\it  Phys. Rev. Lett.} {\bf 80}, 5152
(1998); preprint (to be published).


\bibitem{lassig} M. Lassig, {\it Phys. Rev. Lett.} {\bf 77} 526 (1996).

\bibitem{bhat} S. M. Bhattacharjee, {\it Phys. Rev. Lett.} {\bf 76}
4568 (1996).

\bibitem{mar2} V. I. Marchenko, {\it Sov. Phys. JETP} {\bf 54}, 605
(1981). Note that a factor of 2 should multiply the second line in the
equation at the bottom of the first page of this article.

\bibitem{vzs} V. B. Shenoy, Shiwei Zhang and W. F. Saam, {\it
Phys. Rev. Lett.} {\bf 81} 3475 (1998).

\bibitem{jrs} C. Jayaprakash, C. Rottman, and W. F. Saam, {\it Phys.
Rev.
B} {\bf 30}, 6549 (1984).

\bibitem{deg} P. G. de Gennes, {\it J. Chem. Phys.} {\bf 48} 2257 (1968);
For a general review of fermionic methods, see M. den Nijs, in {\em 
Phase Transitions and Critical Phenomenon}, Vol. 12, edited by
C. Domb and J. L. Lebowitz (Academic, London, 1989).

\bibitem{bk} For a mapping of reconstructed surfaces to the
Hubbard model see L. Balents and M. Kardar, {\it Phys. Rev. B} {\bf 46} 16
031 (1992).

\bibitem{dnv} See for {\em e.g.} C. Duport, P. Nozieres and J. Villain,
{\it Phys. Rev. Lett.} {\bf 74} 134 (1995);
J. Tersoff {\em et.al,} {\it Phys. Rev. Lett.} {\bf 75} 2730
(1995); 
D. Kandel and J. D. Weeks,  {\it  Phys. Rev. Lett.} {\bf 74} 3632
(1996); D-J. Liu and J. D. Weeks, ({\it Cond-Mat}/9803173).



\bibitem{frad} E. Fradkin, {\em Field Theories Of Condensed Matter
Systems},
(Addison-Wesley, Redwood City, California, 1991).

\bibitem{marc} V. I. Marchenko and A. Ya. Parshin, {\it Sov. Phys. JETP}, {\bf
52}
129 (1981).

\bibitem{Noz} P. Nozieres, in {\em Solids Far From
Equilibrium}, edited by C. Godreche (Cambridge University Press,
Cambridge, 1991).

\bibitem{frohn} J. Frohn {\em et.al,} {\it Phys. Rev. Lett.} {\bf 67}
3543 (1991).

\bibitem{heyraud} J. C. Heyraud and J. J. Metois, {\it J. Cryst. Growth} {\bf
50} 571 (1980); {\it Acta Metall.} {\bf 28} 1789 (1980).

\bibitem{red} For a theoretical discussion of the
orgin of attractive step-step interactions see A. C. Redfield and A. Zangwill, {\it Phys. Rev. B} {\bf
46}, 4289 (1992) and references therein.

\bibitem{numr} Details of this method can be found in
W. H. Press {\em et.al.,} {\em Numerical Recipies} (Cambridge
University Press, Cambridge, 1987), page 258.

\bibitem{burk} T. W. Burkhardt and P. Schlottmann, {\it J. Phys. A}
{\bf 26} L501 (1993).

\bibitem{hartree} See Eqs.~(8) and (9) of Ref.~\cite{jrs} for a
derivation of this result.

\bibitem{Suth} B. Sutherland, {\it J. Math. Phys.} {\bf 12} 246, 251 (1971).



\bibitem{GFMC}M.~H.~Kalos, {\it Phys. Rev.} {\bf 128}, 1791 (1962);
D.~M.~Ceperley and M.~H.~Kalos, in {\it Monte Carlo Methods in
Statistical Physics\/}, ed.\ by K.~Binder (Springer-Verlag,
Heidelberg, 1979).

\bibitem{Triv} N. Trivedi, D. M. Ceperley, {\it Phys. Rev. B} {\bf 41}
4552 (1990).

\bibitem{Shi} Shiwei Zhang {\em et.al.,} {\it Phys. Rev. Lett.} {\bf
74} 1500 (1995).


\bibitem{akutsu} Y. Akutsu, N. Akutsu and T. Yamamoto, {\it
Phys. Rev. Lett.}
{\bf 61}, 424 (1988); {\it Phys. Rev. Lett.} {\bf 62}, 2637 (1989).

\bibitem{Saam} W. F. Saam, {\it Phys. Rev. Lett.} {\bf 62}, 2636
(1989).

\bibitem{vand} S. van Dijken {\it et al}, {\it Phys. Rev. B} {\bf 55}, 7864 (1997).

\bibitem{zhu} Jian-hong Zhu {\it et al}, {\it Appl. Phys. Lett.} {\bf 73}, 2438 (1998).

\bibitem{fac2} The formula given in Ref.~\cite{lassig} differs from
this
formula by a factor of 2 that multiplies $g$, because of a factor $1/2$
in the normalization of the kinetic energy part of the Hamiltonian.


\bibitem{cyrus} C.~J.~Umrigar {\it et.~al.\/}, {\it Phys. Rev. Lett.}
{\bf 60} 1719 (1988).

\bibitem{bilinear}K.~J.~Runge, {\it Phys. Rev. B} {\bf 45} 7229 (1992);
Shiwei Zhang and M.~H.~Kalos, 
{\it J.~Stat.~Phys.\/} {\bf 70} 515, 1993.

\end{references}
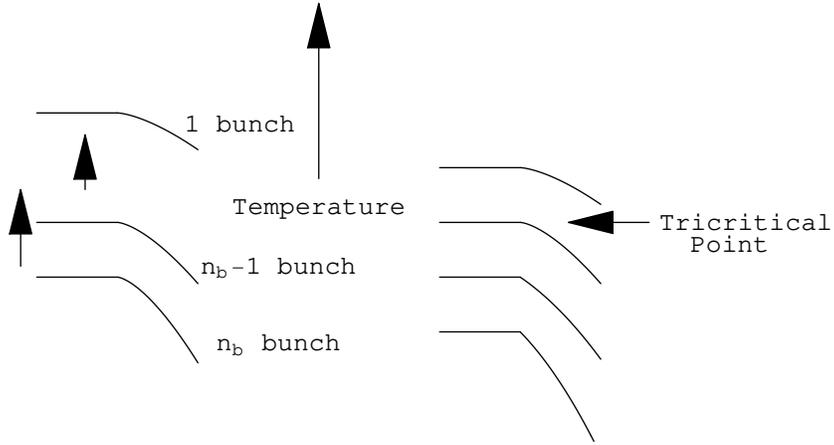
\end{document}